\documentclass[11pt]{article}
\pdfoutput=1 
\usepackage{jheppub}
\usepackage{setspace}
\usepackage{bbm,bm,mathtools,slashed}
\usepackage{multirow}
\usepackage[T1]{fontenc}
\usepackage[nottoc]{tocbibind}
\usepackage[toc,page]{appendix}
\usepackage{youngtab}
\usepackage{physics}

\graphicspath{{./figs/}}

\newcommand{\ri}{\mathrm{i}}

\newcommand{\be}{\begin{equation}}
\newcommand{\ee}{\end{equation}}
\newcommand{\bea}{\begin{eqnarray}}
\newcommand{\eea}{\end{eqnarray}}

\newcommand{\calD}{\mathcal{D}}

\newcommand{\calH}{\mathcal{H}}

\newcommand{\vac}{\Psi_0}

\newlength{\fighskip} \fighskip=2pt
\newlength{\figvskip} \figvskip=3pt

\preprint{KEK-TH-2533, J-PARC-TH-0291, RIKEN-iTHEMS-Report-23} 
\title{$q$ deformed formulation of Hamiltonian SU(3) Yang-Mills theory}

\author[a,b]{Tomoya Hayata,}
\author[b,c,d,e,f]{Yoshimasa Hidaka}

\affiliation[a]{Departments of Physics, Keio University, 4-1-1 Hiyoshi, Kanagawa 223-8521, Japan}
\affiliation[b]{RIKEN iTHEMS, RIKEN, 2-1, Hirosawa, Wako, Saitama 351-0198, Japan}
\affiliation[c]{Theory Center, Institute of Particle and Nuclear Studies, High Energy Accelerator Research Organization (KEK), 1-1 Oho, Tsukuba 305-0801, Japan}
\affiliation[d]{Graduate Institute for Advanced Studies, SOKENDAI, 1-1 Oho, Tsukuba 305-0801, Japan}
\affiliation[e]{nternational Center for Quantum-field Measurement Systems for Studies of the Universe and Particles (QUP), KEK, 1-1 Oho, Tsukuba 305-0801, Japan}
\affiliation[f]{Department of Physics, Faculty of Science, University of Tokyo, 7-3-1 Hongo Bunkyo-ku Tokyo 113-0033, Japan}

\emailAdd{hayata@keio.jp}
\emailAdd{hidaka@post.kek.jp}

\abstract{
We study $\mathrm{SU}(3)$ Yang-Mills theory in $(2+1)$ dimensions based on networks of Wilson lines. 
With the help of the $q$ deformation, networks respect the (discretized) $\mathrm{SU}(3)$ gauge symmetry as a quantum group, i.e., $\mathrm{SU}(3)_k$, and may enable implementations of $\mathrm{SU}(3)$ Yang-Mills theory in quantum and classical algorithms by referring to those of the stringnet model.
As a demonstration, we perform a mean-field computation of the groundstate of $\mathrm{SU}(3)_k$ Yang-Mills theory, which is in good agreement with the conventional Monte Carlo simulation by taking sufficiently large $k$. 
The variational ansatz of the mean-field computation can be represented by the tensor networks called infinite projected entangled pair states.
The success of the mean-field computation indicates that the essential features of Yang-Mills theory are well described by tensor networks, so that they may be useful in numerical simulations of Yang-Mills theory.
}

\begin{document}

\maketitle

\section{Introduction}

Quantum chromodynamics (QCD) is a nonabelian gauge theory describing the strong interaction between quarks mediated by gluons~\cite{Gross:2022hyw}.
One of the most important properties of QCD is the so-called asymptotic freedom: The interaction between quarks and gluons becomes weak and the perturbative computations work remarkably well at short distance such as deep inelastic scattering, while it becomes strong at long distance leading to the confinement of quarks and gluons inside hadrons. 
To understand the physics of confinement and describe hadrons from QCD, we need to  deal with QCD nonperturbatively.
To this end, lattice QCD has been developed, which provides us the most established and powerful computational methods of quantum field theories based on Monte Carlo techniques.
However, some problems remain unsolved such as the QCD phase diagram or equation of state at low temperature and high density, and real-time dynamics of QCD~\cite{deForcrand:2009zkb,Aarts:2015tyj,Alexandru:2020wrj,Nagata:2021ugx}. In those problems, the importance sampling breaks down due to the complex measure of path integrals, which is the notorious sign problem.

Inspired by the recent developments in quantum technology- and information-based tools such as quantum simulation~\cite{cirac_goals_2012,Georgescu:2013oza} and tensor networks~\cite{Orus:2013kga,Cirac:2020obd} in condensed matter systems, much efforts have been devoted to exploring the potential of those tools in high-energy physics~\cite{Dalmonte:2016alw,Preskill:2018fag}.
Indeed, those tools were found to be effective, e.g., in studying nonequilibrium dynamics of lattice gauge theories in $(1+1)$ dimensions~\cite{Banuls:2019rao,Banuls:2019bmf}.
However, we face many challenges in applications to higher dimensions since we need to tackle with the complex dynamics of gauge fields~\cite{Zohar:2021nyc}.
Even if we limit ourselves to pure Yang-Mills theories, i.e., nonabelian gauge theories without quarks for simplifying the problem as is commonly done in theoretical studies of QCD, implementation of gauge fields on quantum simulators or tensor networks is still a nontrivial task: We need to approximate infinite-dimensional Hilbert space of gauge fields as finite degrees of freedom with keeping gauge invariance manifestly.
Although it is not fully established yet, there is some progress in $\mathrm{SU(2)}$~\cite{Anishetty:2018vod,Raychowdhury:2018tfj,Klco:2019evd,Raychowdhury:2019iki,Cunningham:2020uco,ARahman:2021ktn,Hayata:2021kcp,Gonzalez-Cuadra:2022hxt,Yao:2023pht,Zache:2023dko,Hayata:2023puo,Halimeh:2023wrx}. 
However, only a little is known about $\mathrm{SU(3)}$~\cite{Byrnes:2005qx,Ciavarella:2021nmj,Ciavarella:2021lel} although it is preliminary to quantum or classical simulation of QCD.

In this paper, we formulate a regularized Kogut-Susskind Hamiltonian~\cite{Kogut:1974ag} of $\mathrm{SU(3)}$ lattice Yang-Mills theory in $(2+1)$ dimensions based on networks of Wilson lines, which are known as spin networks for $\mathrm{SU(2)}$~\cite{penrose1971angular,Rovelli:1995ac,Baez:1994hx,Burgio:1999tg,Dittrich:2018dvs}, by generalizing the formulation described in refs.~\cite{Cunningham:2020uco,Zache:2023dko,Hayata:2023puo}.
To approximate continuous space of $\mathrm{SU(3)}$ as finite degrees of freedom, we deform $\mathrm{SU(3)}$ group into $\mathrm{SU}(3)_k$ quantum group, and construct the Kogut-Susskind Hamiltonian of $\mathrm{SU}(3)_k$ Yang-Mills theory.
As its first application, we perform the mean-field computation of the groundstate. 
We study the $k$ dependence of observables to estimate $k$ required for discussing the physics in the $k\rightarrow\infty$ limit.

The remainder of this paper is organized as follows.  
We review algebras of networks and the $q$ deformation in section~\ref{sec:stringnet}. 
The crucial difference between $\mathrm{SU(2)}$, and $\mathrm{SU(3)}$ or $\mathrm{SU}(N)$ [$N>2$] is multiplicity of representations, which is elaborated in section~\ref{sec:stringnet}.
Then, using algebras of networks, we construct the Kogut-Susskind Hamiltonian of $\mathrm{SU}(3)_k$ Yang-Mills theory on a square lattice in section~\ref{sec:KS}.
We give technical details of the mean-field computation, particularly the computation of expectation values based on graphs in section~\ref{sec:tdvp}. Numerical results are shown in section~\ref{sec:numerical}.
We compare the mean-field computation of $\mathrm{SU}(3)_k$ Yang-Mills theory with the Monte Carlo simulation of $\mathrm{SU(3)}$ Yang-Mills theory.
Finally, we briefly discuss future improvements of this work in section~\ref{sec:discussion}. 

\section{Networks and algebras}
\label{sec:stringnet}
\subsection{Algebraic data}
The gauge invariant physical Hilbert space of a lattice gauge theory can be represented using a basis that corresponds to a network of Wilson lines.
In the case of SU(2), such a network is known as a spin network~\cite{penrose1971angular,Rovelli:1995ac,Baez:1994hx,Burgio:1999tg,Dittrich:2018dvs}.
If the gauge group is continuous, the Hilbert space of gauge fields exhibits infinite degrees of freedom even on a finite lattice, and it is necessary to approximate the Hilbert space as finite degrees of freedom for quantum simulations.
We regularize the theory by deforming the $\mathrm{SU}(3)$ group into the $\mathrm{SU}(3)_k$ quantum group.
The original $\mathrm{SU}(3)$ group is recovered by the $k\to\infty$ limit.
The algebra generated by the network of Wilson lines of the quantum group is described by the unitary modular tensor category (UMTC), which is employed as an anyon model~\cite{Kitaev:2005hzj}.
In this section, we briefly summarize the algebra of networks necessary for Hamiltonian formalism, following the conventions in ref.~\cite{Barkeshli:2014cna}. For more comprehensive details, see, e.g., references \cite{Kitaev:2005hzj,Bonderson:2007ci}.

Wilson lines are labeled by the representations of the gauge group, and we represent them by $a,b,c,\cdots$. For $\mathrm{SU}(2)_k$ gauge theory,  these representations are equivalent to the angular momentum $j$.
In $\mathrm{SU}(N)_k$, they correspond to the highest weight vectors or the so-called Dynkin labels.
For a given representation $a$, there exists an anti-representation, which we write as $\bar{a}$. 
They are graphically represented by directed lines:
 \begin{equation}
  \parbox{.6cm}{\includegraphics[scale=0.3]{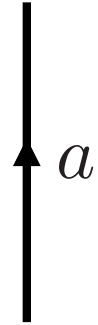}}
  =\
  \parbox{.6cm}{\includegraphics[scale=0.3]{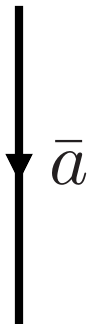}}.
 \end{equation}
 Wilson lines can be composed. The product of Wilson lines is decomposed into irreducible representations.
 This decomposition is represented by the fusion rule:
\begin{equation}
  a\times b = \sum_{c}N_{ab}^{c}c,
\end{equation}
which satisfies the associativity,
\begin{equation}
\sum_e  N_{ab}^{e} N_{ec}^{d} = \sum_f N_{af}^{d} N_{bc}^{f},
\end{equation}
and $N_{ab}^c=N_{ba}^c=N_{b\bar{c}}^{\bar{a}}=N_{\bar{a}\bar{b}}^{\bar{c}}$.
$N_{ab}^{c}$ is the multiplicity of the composition of representations, which is a nonnegative integer.
For example, in $\mathrm{SU}(3)$ group, the product of the adjoint representation $\vb{8}$,
$\vb{8}\otimes \vb{8}$, is decomposed into $\vb{1}\oplus\vb{8}\oplus\vb{8}\oplus\overline{\vb{10}}\oplus\vb{10}\oplus\vb{27}$, which implies 
$N_{\vb{8}\vb{8}}^{\vb{1}}=N_{\vb{8}\vb{8}}^{\vb{10}}=N_{\vb{8}\vb{8}}^{\overline{\vb{10}}}=N_{\vb{8}\vb{8}}^{{\vb{27}}}=1$ and $N_{\vb{8}\vb{8}}^{\vb{8}}=2$.
Here, we have used the dimension of the representations as a label of representations. If we use Dynkin labels $(p,q)$, the correspondences are $\vb{1}=(0,0)$, $\vb{8}=(1,1)$, $\vb{10}=(2,0)$, $\overline{\vb{10}}=(0,2)$, and ${\vb{27}}=(2,2)$.

When $\vb{N}_a$ is regarded as a matrix whose matrix components are given by $[\vb{N}_a]_b^c=N_{ab}^c$, the largest eigenvalue of $\vb{N}_a$ is called the quantum dimension $d_a$, which is generically not integers.
From $N_{ab}^c=N_{\bar{a}\bar{b}}^{\bar{c}}$, the quantum dimension of the anti-representation $\bar{a}$ is equal to that of the representation $a$, i.e., $d_{\bar{a}}=d_a$.
At $k\to \infty$, they are reduced to the dimensions of the ordinary representation matrices, e.g., $d_{(1,0)}=d_{(0,1)}=3$, $d_{(1,1)}=8$, etc. The complete expression for $\mathrm{SU}(3)_k$ is shown in section~\ref{sec:su3_k}.

A network of Wilson lines may have a junction, which is labeled by an additional quantum number $\mu$ to distinguish states with multiplicity when $ N_{ab}^c\geq2$:
\begin{align}
  \parbox{1.5cm}{\includegraphics[scale=0.3]{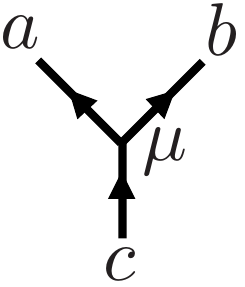}},\quad
  \parbox{1.5cm}{\includegraphics[scale=0.3]{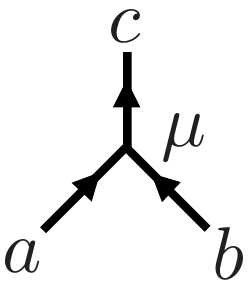}},
\end{align}
where the vertex label runs $\mu\in\{1,2,\cdots, N_{ab}^c\}$.
Each junction must satisfy $N_{ab}^c>0$.

To construct the algebra of a network of Wilson lines, we need information about the local composition, decomposition, and fusion transformations of Wilson lines at junctions. This information can be specifically derived from the composition and decomposition of representations of (quantum) groups.
This algebra characterizes the algebra of anyons in a topological phase and possesses a topological nature, implying that the anyons  or equivalently Wilson lines can be freely deformed as long as they do not intersect in spacetime.
The graphical representation of the topological deformation rules can be summarized as follows:
\begin{align}
  \parbox{1.3cm}{\includegraphics[scale=0.3]{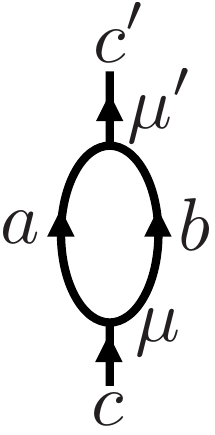}}
&= \delta_c^{c'}\delta_{\mu}^{\mu'}\sqrt{\frac{d_ad_b}{d_c}}
\ \parbox{1.5cm}{\includegraphics[scale=0.3]{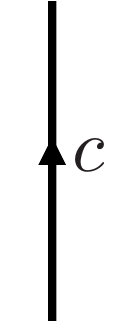}},\label{eq:stacking}\\
  \ \parbox{1.3cm}{\includegraphics[scale=0.3]{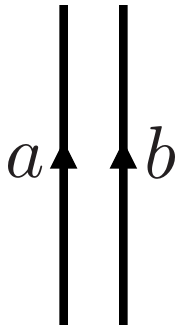}}
&= \sum_{c,\mu}\sqrt{\frac{d_c}{d_ad_b}}
\ \parbox{1.5cm}{\includegraphics[scale=0.3]{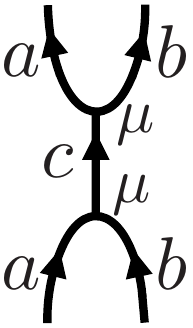}}\label{eq:partition},\\
  \ \parbox{2.3cm}{\includegraphics[scale=0.3]{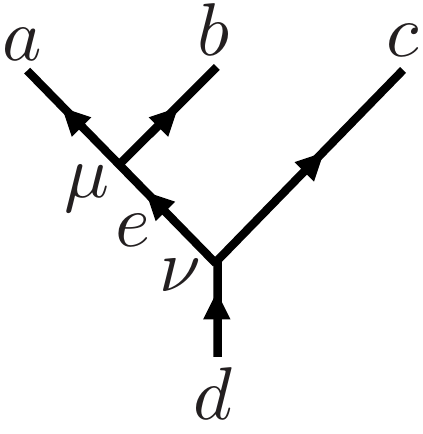}}
&= \sum_{f,\rho,\sigma} [F^{abc}_d]_{(e,\mu,\nu)(f,\rho,\sigma)}
\ \parbox{2.3cm}{\includegraphics[scale=0.3]{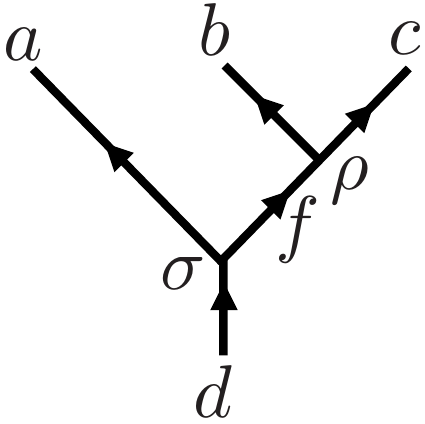}}.\label{eq:Fmove}
\end{align}
Here, $[F^{abc}_d]_{(e,\mu,\nu)(f,\rho,\sigma)}$ are called the $F$-symbols. They are unitary matrices, i.e.,
\begin{equation}
  [(F^{abc}_d)^{-1}]_{(e,\mu,\nu)(f,\rho,\sigma)}
  =[(F^{abc}_d)^{\dag}]_{(e,\mu,\nu)(f,\rho,\sigma)}
  =[F^{abc}_d]^{*}_{(f,\rho,\sigma)(e,\mu,\nu)},
\end{equation}
and satisfy the pentagon equation
\begin{equation}
  \begin{split}
    &\sum_{\delta}[F_e^{fcd}]_{(g,\beta,\gamma)(l,\nu,\delta)} [F_e^{abl}]_{(f,\alpha,\delta)(k,\mu,\lambda)}\\
    &\quad= \sum_{h,\sigma,\psi,\rho} [F^{abc}_g]_{(f,\alpha,\beta)(h,\psi,\sigma)} [F_{e}^{ahd}]_{(g,\sigma,\gamma)(k,\rho,\lambda)} [F^{bcd}_{k}]_{(h,\psi,\rho)(l,\nu,\mu)}.
  \end{split}
\end{equation}
In the UMTC, there is also a braiding exchange represented by
\begin{align}
  \parbox{1.3cm}{\includegraphics[scale=0.3]{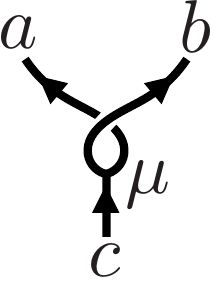}}
=
  \sum_{\nu}[R_c^{ab}]_{\mu\nu}\ \parbox{1.3cm}{\includegraphics[scale=0.3]{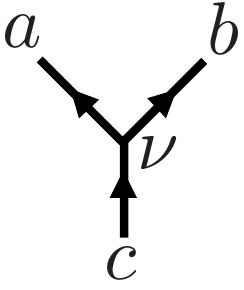}},
\end{align}
which is not used in our paper.
$[R^{ab}]_{\mu\nu}$ must be compatible to the $F$-symbols, and the compatibility condition is described by Hexagon equations.

Let us use the deformation rules to derive some relations.
A Wilson loop with representation $a$ is graphically represented as
\begin{align}
  \tr U_a =\  \parbox{2.cm}{\includegraphics[scale=0.3]{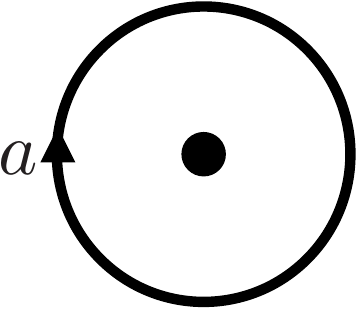}}.
\end{align}
Here, we introduced a defect at the center of the loop to prevent it from collapsing when contracted.
When the defect is absent, it reduces to
\begin{align}\label{eq:loop_deformation}
  \parbox{2.cm}{\includegraphics[scale=0.3]{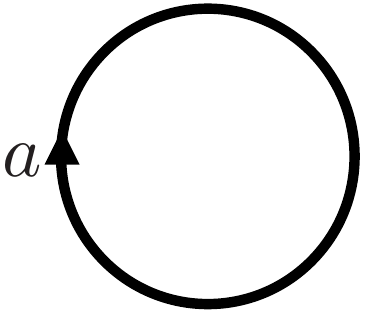}} = d_a,
\end{align}
by choosing $b=\bar{a}$ and $c=0$ in eq.~\eqref{eq:stacking}.
Two Wilson loops with representations $a$ and $b$ can be deformed using eqs~\eqref{eq:stacking} and \eqref{eq:partition} to 
\begin{align}
\label{eq:wilsonloop_fusion_graph}
  \parbox{2.cm}{\includegraphics[scale=0.3]{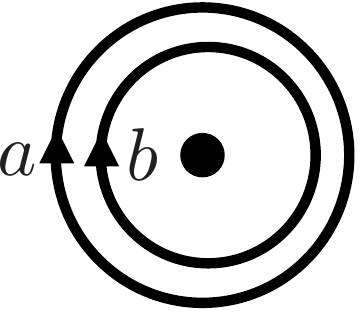}}
  &=\sum_{\mu,c}\frac{\sqrt{d_c}}{\sqrt{d_ad_b}}\ \parbox{2.2cm}{\includegraphics[scale=0.3]{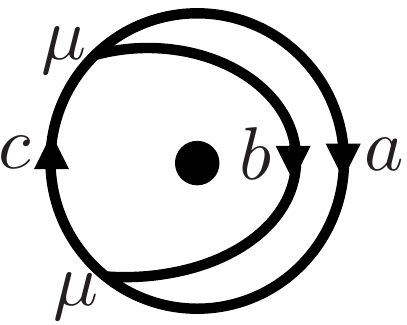}}
  =\sum_{\mu,c}\frac{\sqrt{d_c}}{\sqrt{d_ad_b}}\  \parbox{2.2cm}{\includegraphics[scale=0.3]{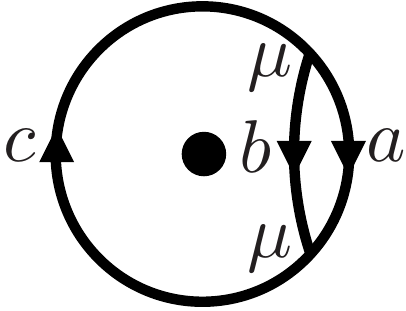}}\notag\\
  &=\sum_{c,\mu}\frac{\sqrt{d_c}}{\sqrt{d_ad_b}}\frac{\sqrt{d_{\bar{b}} d_{\bar{a}}}}{\sqrt{d_{\bar{c}}}}\  \parbox{2.cm}{\includegraphics[scale=0.3]{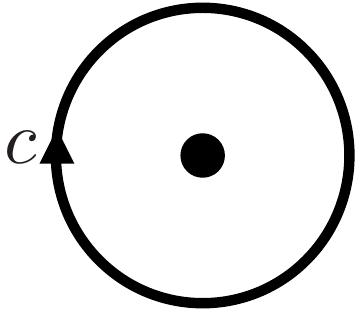}}
  =\sum_{c}N_{ab}^c\  \parbox{2.cm}{\includegraphics[scale=0.3]{figs/loopc.pdf}}.
\end{align}
In the last equality, we used $d_{\bar{a}}=d_{a}$, etc.
This leads to the fusion rule of Wilson loops:
\begin{equation}
\label{eq:wilsonloop_fusion}
  \tr U_a \tr U_b = \sum_{c}N_{ab}^c\tr U_c.
\end{equation}
Furthermore, using eq.~\eqref{eq:loop_deformation}, we obtain
\begin{equation}\label{eq:dadb}
  d_{a}d_{b} =\sum_{c}N_{ab}^c d_c.
\end{equation}
Let us derive the inverse relation of eq.~\eqref{eq:dadb}.
Noting $d_{\bar{b}}=d_{b}$, $d_{\bar{c}}=d_{c}$, and $N_{a\bar{b}}^{\bar{c}}=N_{ac}^b$, we obtain
\begin{equation}
  d_{a}d_{b} =\sum_{c}N_{ac}^{b} d_c.
\end{equation}
Multiplying both sides by $d_a$ and summing over $a$ yield
the inverse relation of eq.~\eqref{eq:dadb},
\begin{equation}\label{eq:inverser_dadb}
  \mathcal{D}^2 d_{b} =\sum_{a,c}N_{ac}^{b} d_a d_c.
\end{equation}
where $\mathcal{D}$ is the total quantum dimension defined by 
\begin{equation}
  \mathcal{D} \coloneqq \sqrt{\sum_{a}d_a^2}.
\end{equation}
In a gauge theory, of course, the Wilson lines are generally not topological. However, as will be seen later, these rules play a crucial role in computing the action of the operators on a physical state.

\subsection{\texorpdfstring{$\mathrm{SU}(3)_k$}{SU(3)k} quantum group}\label{sec:su3_k}
We are interested in $\mathrm{SU(3)}_k$ in our numerical calculations; let us summarize the mathematical objects used in section~\ref{sec:tdvp} such as the fusion multiplicities, second order Casimir invariants, and quantum dimensions for $\mathrm{SU}(3)_k$.
For $\mathrm{SU}(3)_k$, an irreducible representation is represented by Dynkin labels, $a=(p_a,q_a)$, where
$p_a$ and $q_a$ are the number of single and double box columns in the Young tableaux, respectively.

The fusion multiplicities are given as~\cite{Begin:1992rt}
\begin{equation}
  N_{ab}^{c} =(k_0^{\max}-k_0^{\min}+1)\delta_{ab}^c,
\end{equation}
where 
\begin{align}
  k_0^{\min}&=\max(p_a+q_a,p_b+q_b,p_c+q_c,\mathcal{A}-\min(p_a,p_b,q_c),\mathcal{B}-\min(q_a,q_b,p_c)),\\
  k_0^{\max}&=\min(\mathcal{A},\mathcal{B}),\\
  \mathcal{A}&=\frac{1}{3}\qty(2(p_a+p_b+q_c)+q_a+q_b+p_c),\\
  \mathcal{B}&=\frac{1}{3}\qty(p_a+p_b+q_c+2(q_a+q_b+p_c)),\\
  \delta_{ab}^c&=\begin{cases}
    1\quad \text{if $k_0^{\max}>k_0^{\min}$ and $\mathcal{A}, \mathcal{B}\in\mathbb{Z}_+$}\\
    0\quad\text{otherwise}
  \end{cases}.
\end{align}
Here, $\mathbb{Z}_+$ represents the set of non-negative integers.

Using the $q$-deformation parameter,
\begin{equation}
  q = \exp(\ri\frac{2\pi}{k+3}),
\end{equation}
the so-called $q$-number is defined as
\begin{equation}
  [n]\coloneqq  \frac{q^{\frac{n}{2}}-q^{-\frac{n}{2}}}{q^{\frac{1}{2}}-q^{-\frac{1}{2}}}=\frac{\sin\qty(\frac{\pi}{k+3}n)}{\sin\qty(\frac{\pi}{k+3})}.
  \label{eq:[n]}
\end{equation}
Note that the $q$-deformation parameter `$q$' is different from $q$ in the Dynkin labels $(p,q)$.
In the following, $q$ appears only as the Dynkin labels, so there will be no confusion.

The second order Casimir invariant $C_2(a)$ is given as~\cite{Bonatsos:1999xj}\footnote{We choose that the normalization factor of $C_2(a)$ is half of the value used in ref.~\cite{Bonatsos:1999xj}.}.
\begin{equation}\label{eq:C2}
  C_2(a)  = \frac{1}{2}\qty(\left[\frac{p_a}{3}-\frac{q_a}{3}\right]^2+\left[\frac{2p_a}{3}+\frac{q_a}{3}+1\right]^2+\left[\frac{p_a}{3}+\frac{2q_a}{3}+1\right]^2-2).
\end{equation}
Similarly, the quantum dimension is given by~\cite{Coquereaux:2005hu}
\begin{equation}
  d_{a} =\frac{1}{[2]}[p_a+1][q_a+1][p_a+q_a+2].
  \label{eq:dpq}
\end{equation}
To the best of our knowledge, the general compact form of the $F$-symbols is not known for $\mathrm{SU}(3)_k$.
Some special cases for a small $k$ were studied in ref.~\cite{Ardonne2010ClebschGordanA6}.
As discussed in section~\ref{sec:tdvp}, the method we use does not require a specific form of the $F$-symbols.

\section{\texorpdfstring{$\mathrm{SU}(3)_k$}{SU(3)k} Yang Mills theory on a square lattice}
\label{sec:KS}
We consider $\mathrm{SU}(3)_k$ Yang Mills theory on a square lattice in $(2+1)$ dimensions.
Square lattices with four-point vertices are not represented in the algebra of the previous section.
In order to treat them, we deform the square lattice into a honeycomb lattice with three-point vertices by inserting auxiliary edges~\cite{Hayata:2023puo}:
\begin{equation}
  \parbox{2.3cm}{\includegraphics[scale=0.3]{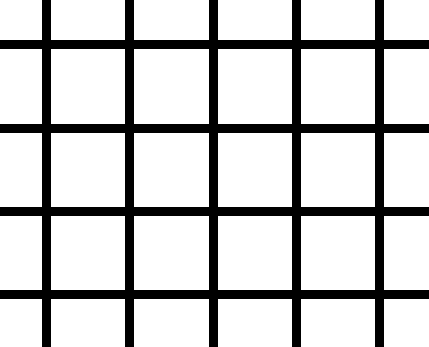}}
  \Rightarrow\
  \parbox{2.3cm}{\includegraphics[scale=0.25]{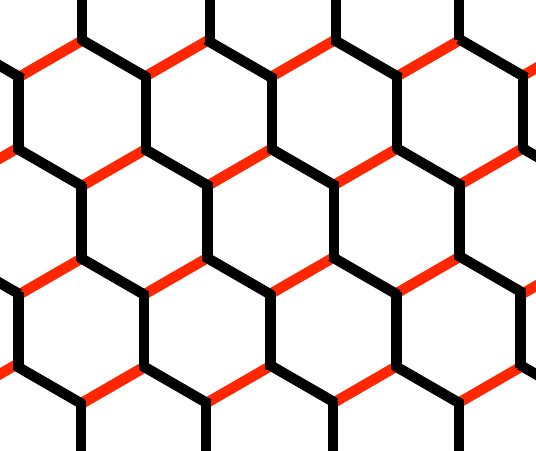}}\quad,
  \label{eq:deformation}
\end{equation}
where the red lines represent the auxiliary edges.
Note that this deformation is equivalent to the original square lattice as long as no electric field operator acts on the auxiliary edges.

This deformation~\eqref{eq:deformation} allows the physical state on the lattice to be represented using Wilson-line networks in the previous section.
A basis of the physical Hilbert space is labeled by the representation and multiplicity on edges and vertices, respectively:
\begin{equation}\label{eq:state_ket_vec}
  \ket{\vb*{a};\vb*{\mu}}=\prod_{e\in \mathcal{E}}\ket{a_e}
  \prod_{v\in \mathcal{V}}\ket{\mu_v},
\end{equation}
which satisfies the orthogonal relation:
\begin{equation}
  \bra{\vb*{a}';\vb*{\mu}'}\ket{\vb*{a};\vb*{\mu}}=\delta_{\vb*{a}',\vb*{a}}\delta_{\vb*{\mu}',\vb*{\mu}}=
  \prod_{e\in \mathcal{E}}\delta_{a'_{e},a_{e}}
  \prod_{v\in \mathcal{V}}\delta_{\mu'_v,\mu_v}.
\end{equation}
Here, $\mathcal{E}$ and $\mathcal{V}$ are the set of edges and vertices.
Note that the incoming $a$, $b$, and outgoing $c$ edges connected to a vertex cannot be arbitrary and must satisfy $N_{ab}^c>0$.
The state $\ket{\vb*{a};\vb*{\mu}}$ is generated by applying a network of Wilson line operators to $\ket{\vb*{0};\vb*{0}}$.
A state around a plaquette is graphically represented by
\begin{equation}
  \prod_{i=1}^6\ket{a_i}\ket{c_i}\ket{\mu_i}=\ \parbox{2.5cm}{\includegraphics[scale=0.212]{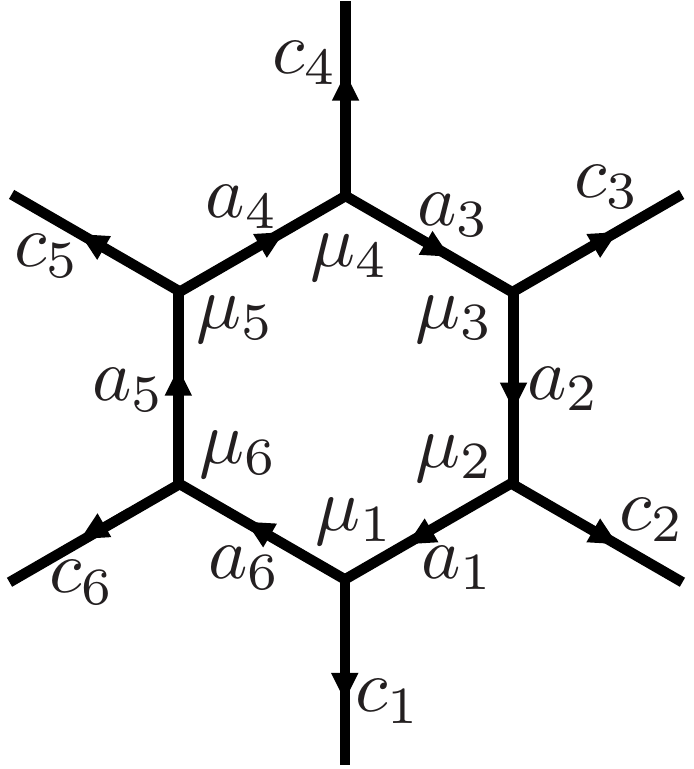}}\quad\qquad.
  \label{eq:network}
\end{equation}

The Kogut-Susskind Hamiltonian of $\mathrm{SU}(3)_k$ Yang-Mills theory has the form,
\begin{equation}
  H = \frac{1}{2}\sum_{e\in {\tilde{\mathcal{E}}}} E_i^2(e) - \frac{K}{2} \sum_{f\in \mathcal{F}}\qty( \tr U_{(1,0)} (f)+\tr U_{(0,1)} (f) ).
  \label{eq:Hamiltonian}
\end{equation}
Here, $\tilde{\mathcal{E}}\subset\mathcal{E}$ is the set of blacked edges in eq.~\eqref{eq:deformation}, and $\mathcal{F}$ is the set of hexagonal plaquettes in eq.~\eqref{eq:deformation}.
$E_i(e)$ is the electric field, i.e., generator of $\mathrm{SU}(3)_k$, and $\tr U_{(1,0)} (f)$ is the Wilson loop that circles the edges of the plaquette $f$ clockwise.
Note that the coupling strength $K$ and conventional coupling constant $g$ can be related as $K=1/g^4$ in the lattice unit.

Let us consider the action of the Hamiltonian on a state.
The action of the electric field term in the Hamiltonian, i.e., $E_i^2$ term, is graphically represented as
\begin{equation}
  E_i^2 \ \parbox{.8cm}{\includegraphics[scale=0.3]{figs/a.pdf}}
  = C_2(a)\
  \parbox{.8cm}{\includegraphics[scale=0.3]{figs/a.pdf}}.
  \label{eq:action_electric_field}
\end{equation}
Here, $C_2(a)$ is the second order Casimir invariant given in Eq.~\eqref{eq:C2}\footnote{There may be a choice of the action of the electric fields. We employ the $q$-deformed Casimir invariant to ensure consistency with the quantum group structure. On the other hand, a previous study uses the Casimir invariant without the $q$ deformation~\cite{Zache:2023dko}.}.
On the other hand, the action of a Wilson loop $\tr U_{d}$ 
with $d=(1,0)$ or $(0,1)$ is given as~\cite{Levin:2004mi}
\begin{equation}
  \tr U_{d}\quad
  \parbox{2.7cm}{\includegraphics[scale=0.21]{figs/plaquetteAction0.pdf}}
       =\sum_{\{a'_{i},\mu'_i\}}\prod_{i=1}^{6}
       [F_{a'_{i}}^{c_{i}a_{i-1}d}]_{(a_{i},\mu_i,\mu_{i+1}),({a'}_{i-1},\mu'_{i-1},\mu'_{i})}
       \ \parbox{2.7cm}{\includegraphics[scale=0.21 ]{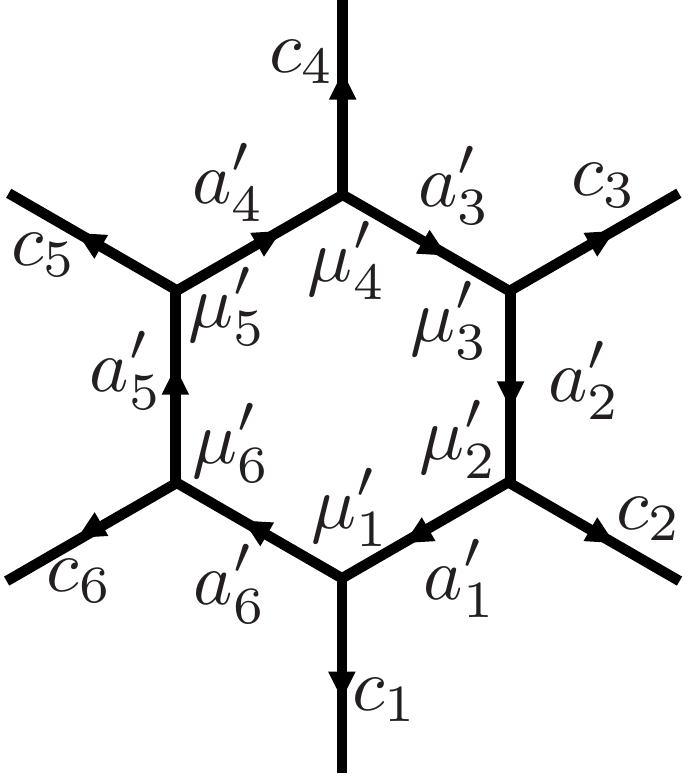}},
         \label{eq:action_trU}
\end{equation}
where $a'_0=a'_6$, $\mu_0=\mu_6$, $\mu'_0=\mu'_6$, and $\mu'_7=\mu'_1$.
We note that eq.~\eqref{eq:action_trU} works for any representation.
Equation~\eqref{eq:action_trU} can be graphically derived by using the topological deformation rules discussed in the previous section.
First, let us put defects in the center of all plaquettes.
For a plaquette, it is graphically represented as
\begin{equation}
  \parbox{2.7cm}{\includegraphics[scale=0.3]{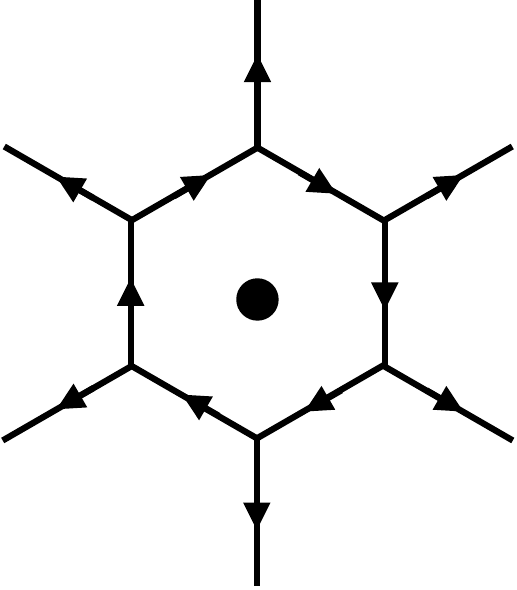}}.
\end{equation}
Here, the indices have been dropped for brevity.
The defect is responsible for making the network non-trivial in its topological deformation.
The action of $\tr U_d$ on a plaquette is expressed by a loop encircling the defect,
\begin{equation}
  \tr U_d \parbox{2.7cm}{\includegraphics[scale=0.3]{figs/loop9.pdf}}
  =\ \parbox{2.7cm}{\includegraphics[scale=0.3]{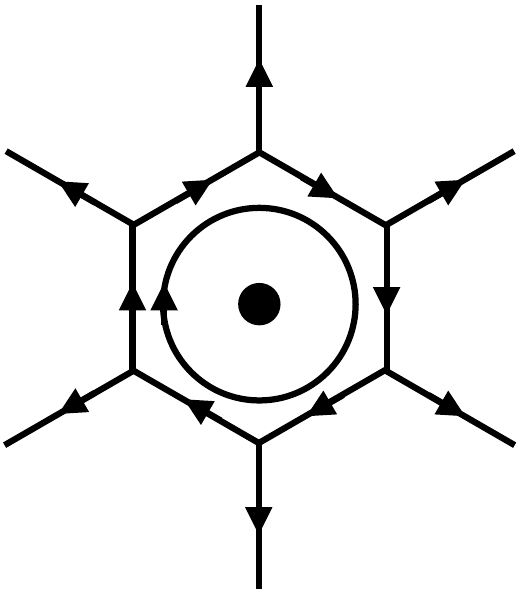}}.
\end{equation}
Using the deformation rules in eqs~\eqref{eq:stacking}-\eqref{eq:Fmove}:
\begin{equation}
  \begin{split}
    &\parbox{2.7cm}{\includegraphics[scale=0.3]{figs/loop1.pdf}}
    \overset{\text{eq.~\eqref{eq:partition}}}{\Longrightarrow}
    \ \parbox{2.7cm}{\includegraphics[scale=0.3]{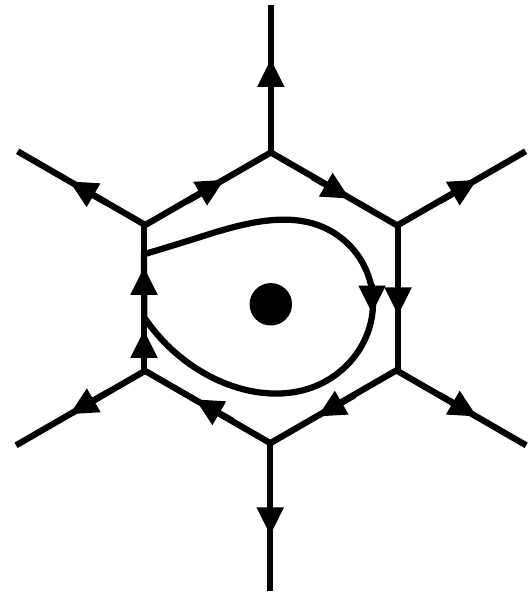}}
    \overset{\text{eq.~\eqref{eq:Fmove}}}{\Longrightarrow}
    \ \parbox{2.7cm}{\includegraphics[scale=0.3]{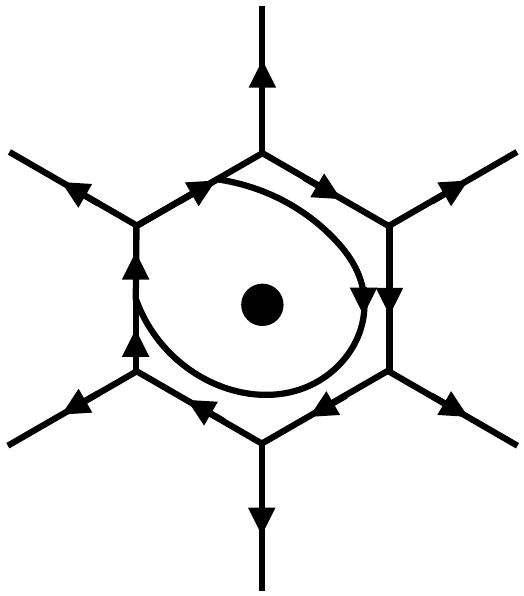}}\\
    &
    \overset{\text{eq.~\eqref{eq:Fmove}}}{\Longrightarrow}
    \ \parbox{2.7cm}{\includegraphics[scale=0.3]{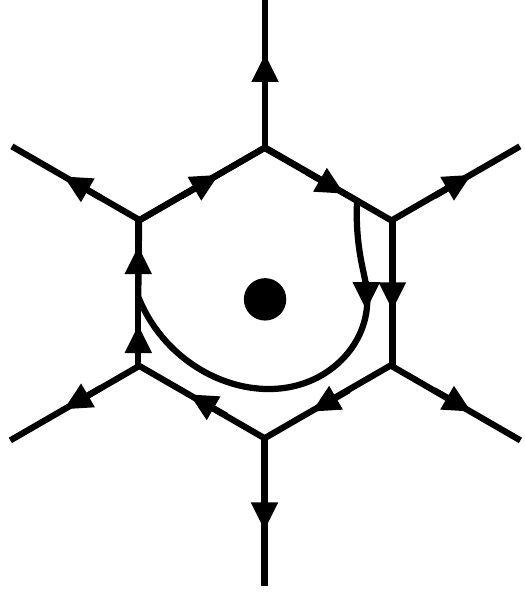}}
    \overset{\text{eq.~\eqref{eq:Fmove}}}{\Longrightarrow}
    \ \parbox{2.7cm}{\includegraphics[scale=0.3]{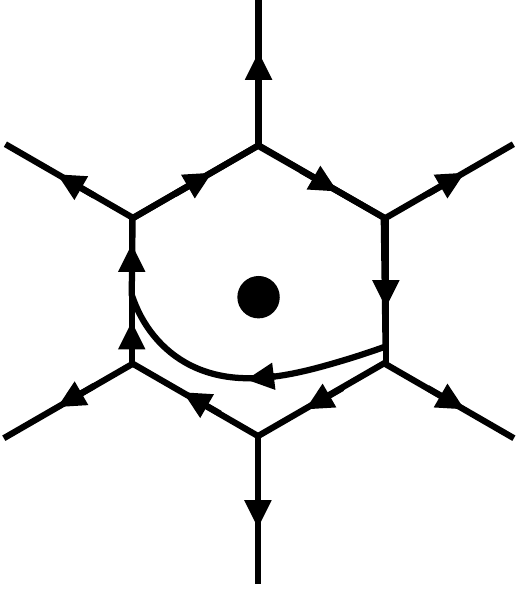}}
    \overset{\text{eq.~\eqref{eq:Fmove}}}{\Longrightarrow}
    \ \parbox{2.7cm}{\includegraphics[scale=0.3]{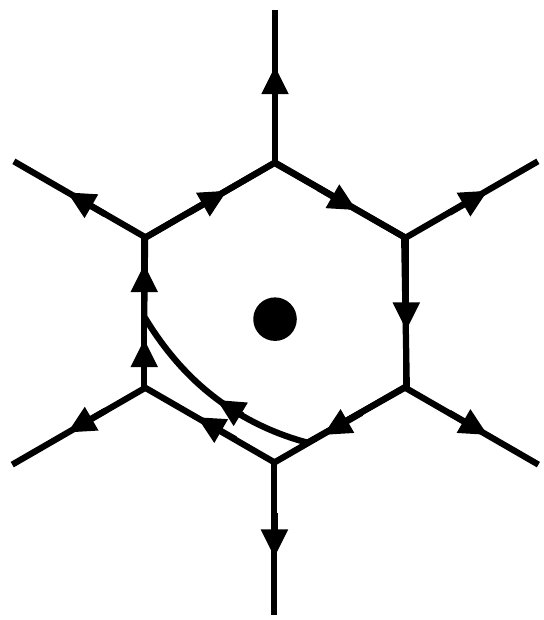}}\\
    &\overset{\text{eq.~\eqref{eq:Fmove}}}{\Longrightarrow}
    \ \parbox{2.7cm}{\includegraphics[scale=0.3]{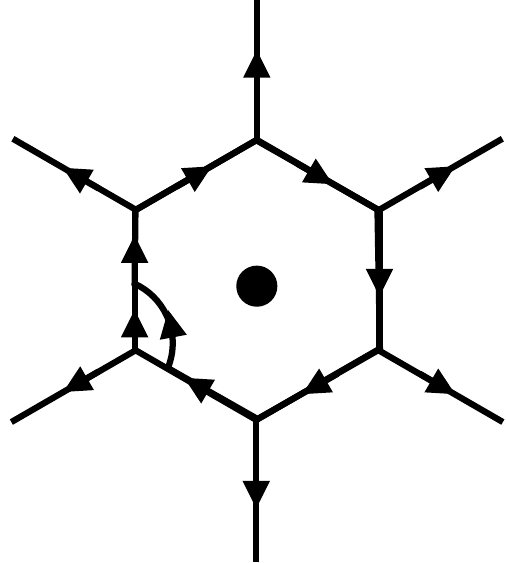}}
    \overset{\text{eq.~\eqref{eq:Fmove}}}{\Longrightarrow}
    \ \parbox{2.7cm}{\includegraphics[scale=0.3]{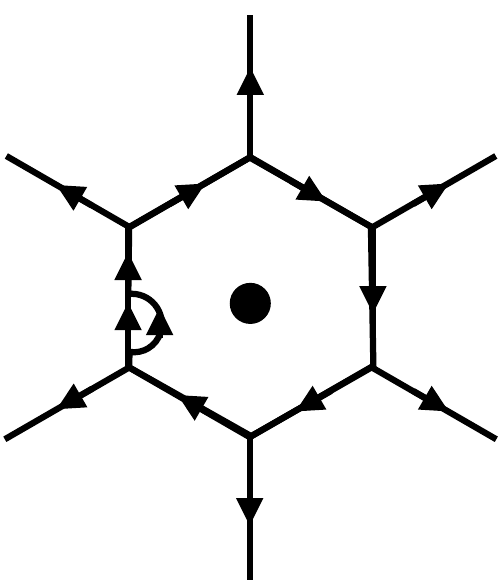}}
    \overset{\text{eq.~\eqref{eq:stacking}}}{\Longrightarrow}
    \ \parbox{2.7cm}{\includegraphics[scale=0.3]{figs/loop9.pdf}},
  \end{split}
\end{equation}
we obtain eq.~\eqref{eq:action_trU}, by tracing the changes of indices.

\section{Computational methods}
\label{sec:tdvp}

We employ a variational ansatz introduced in refs.~\cite{Dusuel:2015sta,Zache:2023dko},
\be
|\Psi\rangle=\prod_{f\in \mathcal{F}} \sum_{a_f}\psi(a_f)\tr U_{a_f}(f)\ket{\vac} ,
\label{eq:wavefunction}
\ee
where $\ket{\vac}=\ket{\vb*{0};\vb*{0}}$, and  variational parameters $\psi(a)$ are normalized as $\sum_a |\psi(a)|^2=1$ such that $\langle\Psi|\Psi\rangle=1$.
We impose open boundary conditions and take the infinite volume limit. Assuming the translational invariance of the groundstate, we employ the same wave function for all plaquettes.
In the following subsections, we explain how we compute the expectation value of an operator in $|\Psi\rangle$ using graph rules introduced in the previous sections, and how we solve the variational problem.

\subsection{Expectation values of observables}
Let us evaluate the expectation value of an operator $\expval{O}\coloneqq\expval{O}{\Psi}$.
As an example, we consider the expectation value of a Wilson loop, $O=\tr U_d(\partial S)$, where $\partial S$ represents the path of the Wilson loop, which is the boundary of an area $S$.
The expectation value is given as
\begin{equation}\label{eq:expvalue_Wilson_loop}
  \begin{split}
    &\expval{\tr U_d(\partial S)}\\
    &= 
    \bra{\vac}\tr U_{d}(\partial S)\qty(\prod_{f\in \mathcal{F}} \sum_{a_f,b_f}\psi^*(a_f)\psi(b_f)\tr U_{\bar{a}_f}(f)\tr U_{b_f}(f))\ket{\vac}\\
    &= 
    \sum_{\{a\},\{b\},\{c\}}\qty(\prod_{f'\in\mathcal{F}}\psi^*(a_{f'})\psi(b_{f'})N_{\bar{a}_{f'}b_{f'}}^{c_{f'}})
    \bra{\vac}\tr U_{d}(\partial S)\prod_{f\in \mathcal{F}} \tr U_{c_f}(f)\ket{\vac},
  \end{split}
\end{equation}
where we introduce a shorthand notation,
\begin{equation}
  \sum_{\{a\}}=\prod_{f\in\mathcal{F}}\sum_{a_f}.
\end{equation}
We repeatedly used the fact that all Wilson loops commute with each other. We also used  
\begin{align}
  \tr U_{b_f}(f)\ket{\vac} =\  \parbox{2.cm}{\includegraphics[scale=0.3]{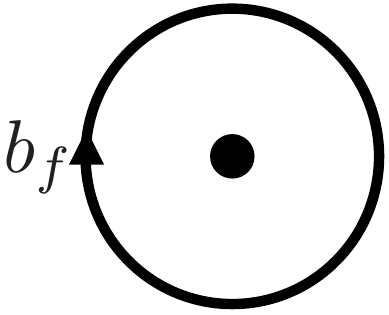}},
\end{align}
and the fusion rule of the Wilson loops in eqs.~\eqref{eq:wilsonloop_fusion_graph} and~\eqref{eq:wilsonloop_fusion} on the same plaquette to obtain the second line.
The nontrivial part of the calculation is now the evaluation of 
\begin{equation}
  \bra{\vac}\tr U_{d}(\partial S)\prod_{f\in \mathcal{F}} \tr U_{c_f}(f)\ket{\vac}.
\end{equation}
To this end, let us consider the state 
\begin{equation}\label{eq:state}
  \tr U_{d}(\partial S)\prod_{f\in \mathcal{F}} \tr U_{c_f}(f)\ket{\vac} ,
\end{equation}
and then apply $\bra{\vac}$ on the state.
We consider a small lattice with boundaries for illustrating the computation, and the state~\eqref{eq:state} is represented as
\begin{equation}
  \begin{split}\label{eq:action_Wilson_loop}
    U_{d}(\partial S)\prod_{f\in \mathcal{F}} \tr U_{c_f}(f)\ket{\vac}=\ \parbox{6.cm}{\includegraphics[scale=0.25]{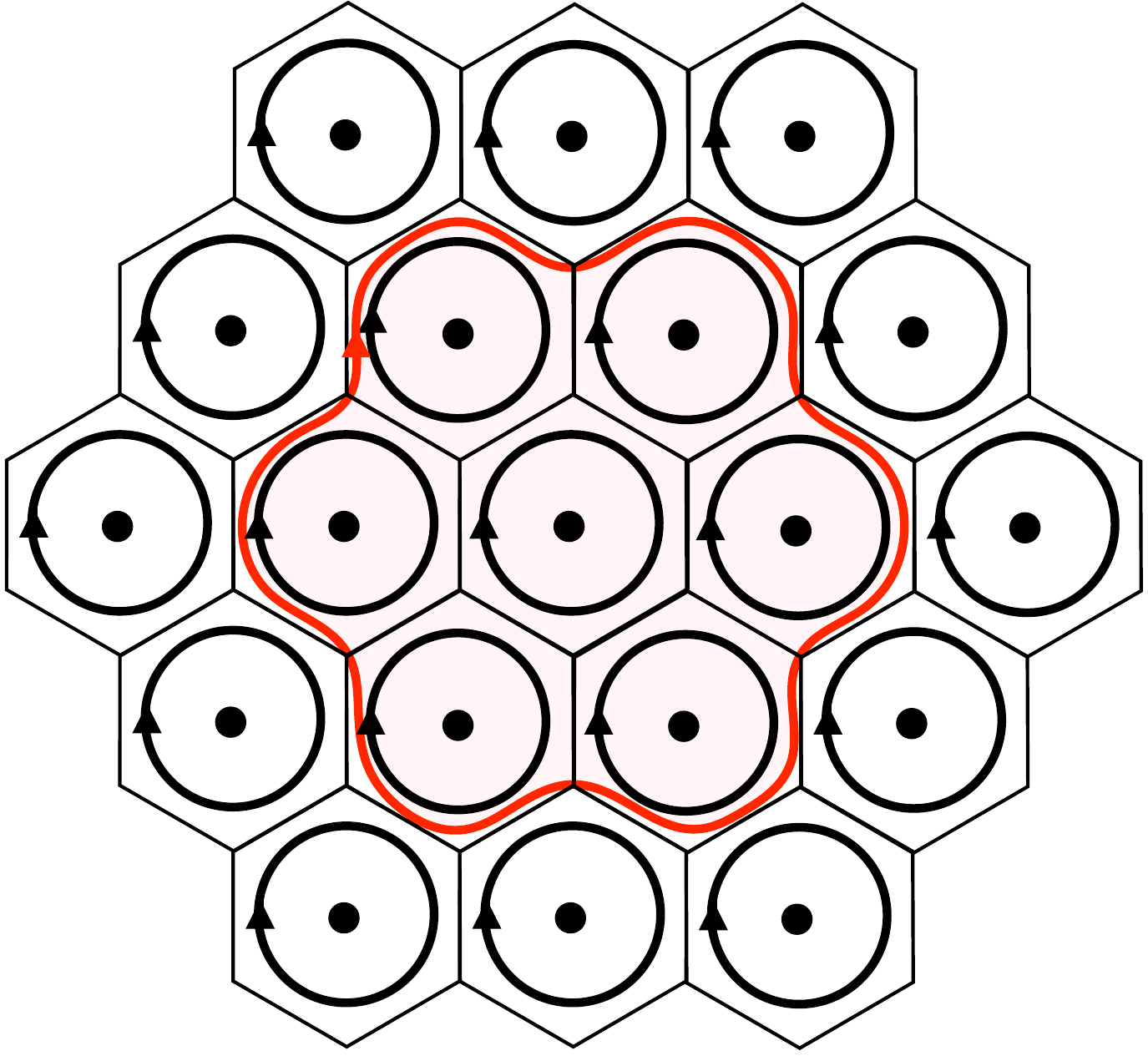}},
  \end{split}
\end{equation}
where the red line is the path of $\partial S$, and 
the inside region shaded by light gray is $S$.
By acting $\bra{\vac}$ on the state, we evaluate the expectation value.
Let us first look at the boundary.
A given boundary plaquette can be deformed as
\begin{equation}
\label{eq:plaquette_boundary}
  \begin{split}
    \parbox{2.2cm}{\includegraphics[scale=0.25]{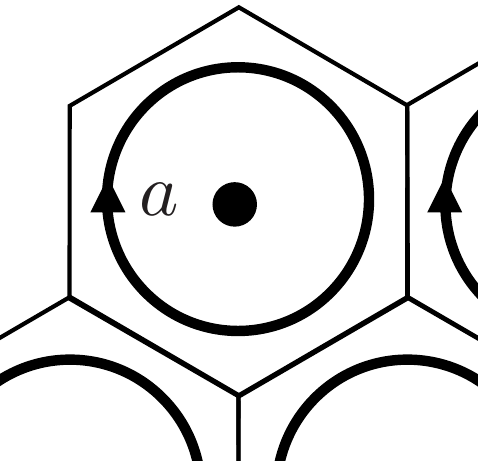}}=\parbox{2.2cm}{\includegraphics[scale=0.25]{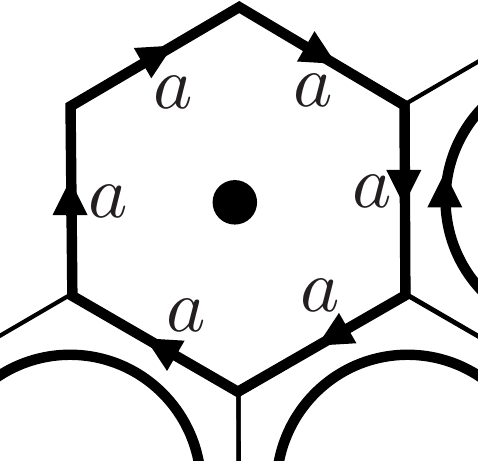}}
  \end{split}.
\end{equation}
When we compute the partial inner product with $\bra{\vac}$ for three edges on the boundary in eq.~\eqref{eq:plaquette_boundary}, only states with trivial representation survive,
since $\bra{0}\ket{a}=\delta_{a}^0$. Thus, we can replace the plaquette operator by
\begin{equation}
\label{eq:reduction}
  \begin{split}
    \parbox{2.2cm}{\includegraphics[scale=0.25]{figs/boundary1.pdf}}= \ \parbox{2.2cm}{\includegraphics[scale=0.25]{figs/boundary2.pdf}}
    \to  \delta_{a}^0\ \parbox{2.2cm}{\includegraphics[scale=0.25]{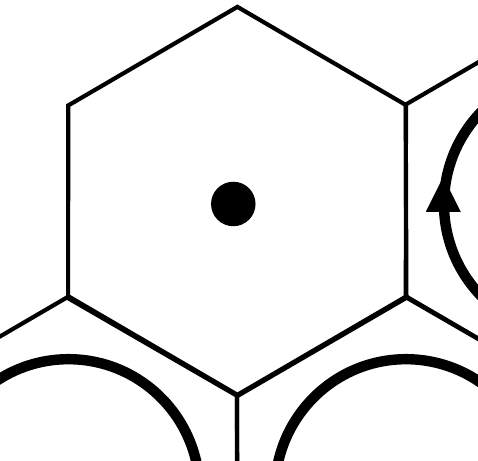}}.
  \end{split}
\end{equation}
The contribution of this plaquette to the expectation value~\eqref{eq:expvalue_Wilson_loop} becomes trivial:
\begin{equation}
  \sum_{a,b,c}\psi^*(a)\psi(b)N_{\bar{a}b}^{c}\delta_{c}^0
  =\sum_{a,b}\psi^*(a)\psi(b)\delta^{a}_b =1,
\end{equation}
where we employ $N_{\bar{a}b}^0=\delta^{a}_b$.
Repeating this procedure until we reach $\partial S$, eq.~\eqref{eq:action_Wilson_loop} becomes
\begin{equation}
  \begin{split}
    \parbox{6.cm}{\includegraphics[scale=0.25]{figs/large_Wilson_loop1.pdf}}
    \to \
    \parbox{6.cm}{\includegraphics[scale=0.25]{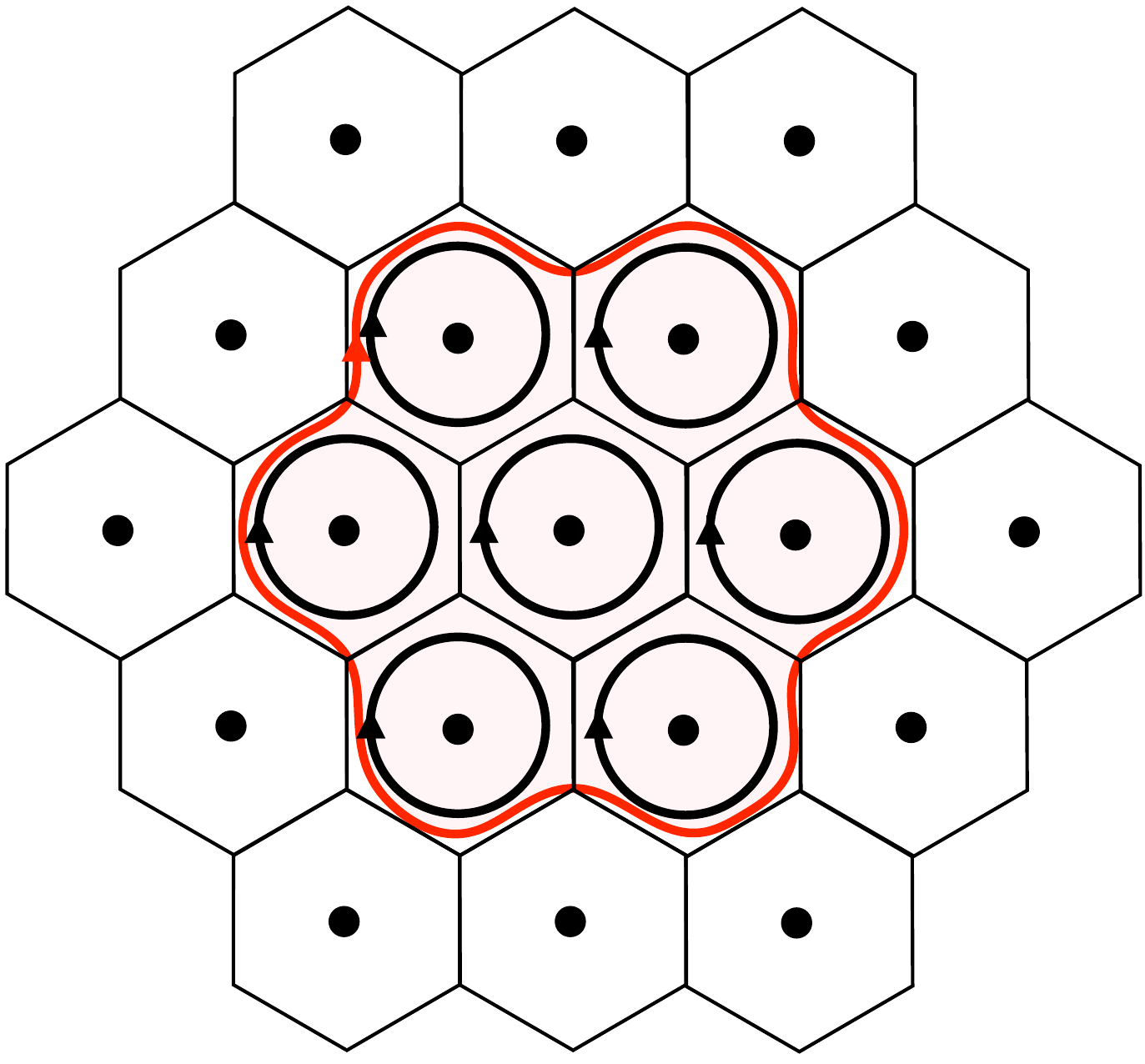}}.
  \end{split}
\end{equation}
Where the Wilson loops overlap, we can use eq.~\eqref{eq:partition} to evaluate
\begin{equation}
\label{eq:reduction2}
  \parbox{3.cm}{\includegraphics[scale=0.3]{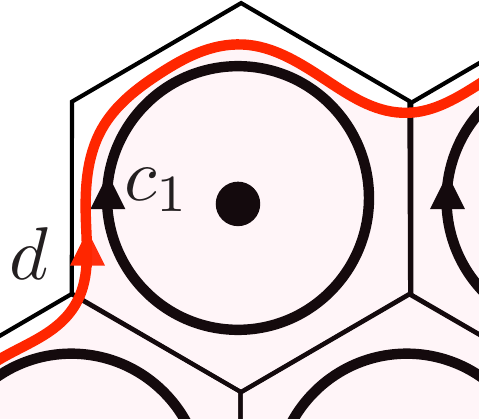}}
  =\sum_{c,\mu}\frac{\sqrt{d_c}}{\sqrt{d_{{d}}d_{c_1}}} \
  \parbox{3.0cm}{\includegraphics[scale=0.3]{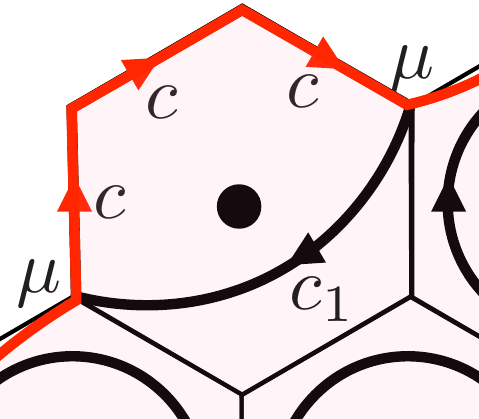}}
  \to
  \frac{\delta^{\bar{d}}_{c_1}}{d_d} \
  \parbox{3.0cm}{\includegraphics[scale=0.3]{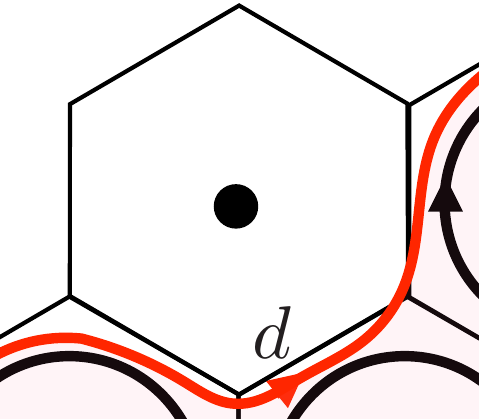}}.
\end{equation}
Here we used the same reduction as eq.~\eqref{eq:reduction}.
Repeating this procedure until the red line shrinks to a point, we obtain
\begin{equation}
  \begin{split}
    \parbox{4.5cm}{\includegraphics[scale=0.25]{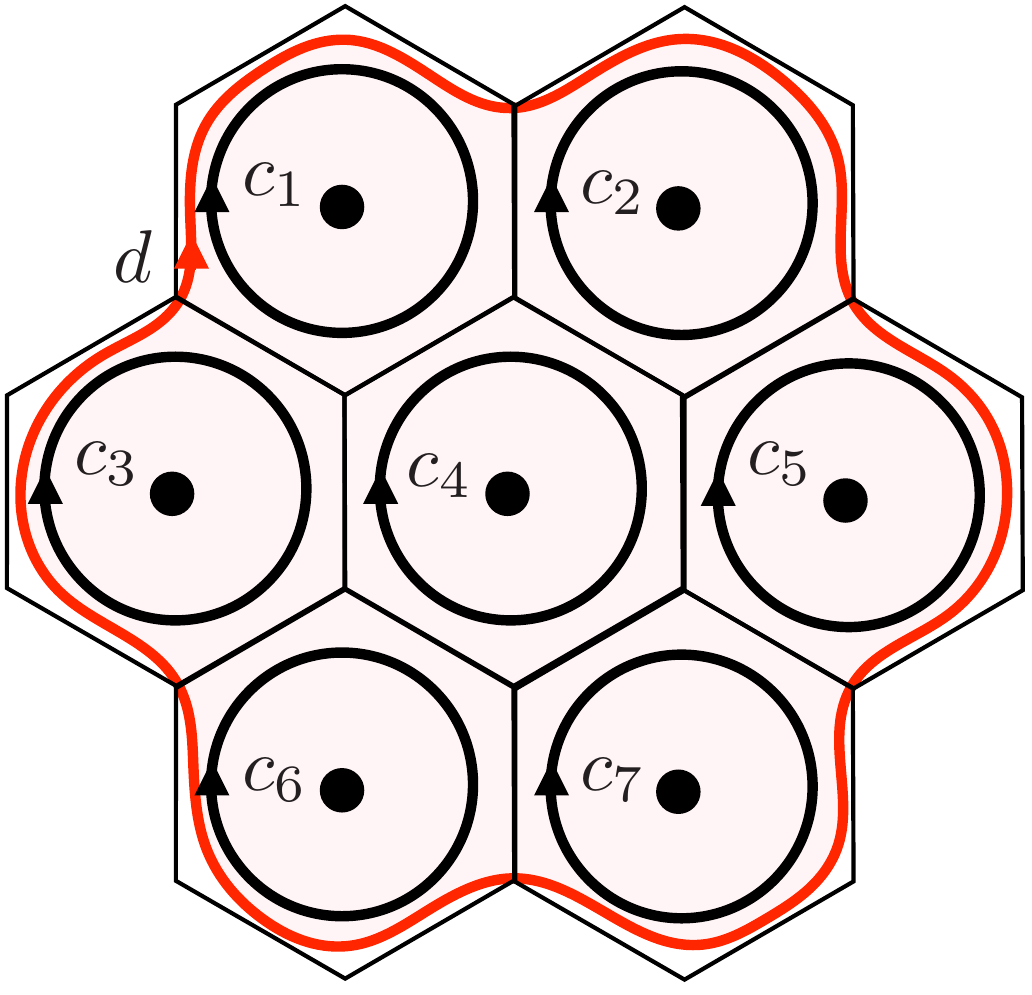}}
    &\to
    \qty(\prod_{f\in S}\frac{\delta^{\bar{d}}_{c_i}}{d_{d}}) \   \parbox{4.5cm}{\includegraphics[scale=0.25]{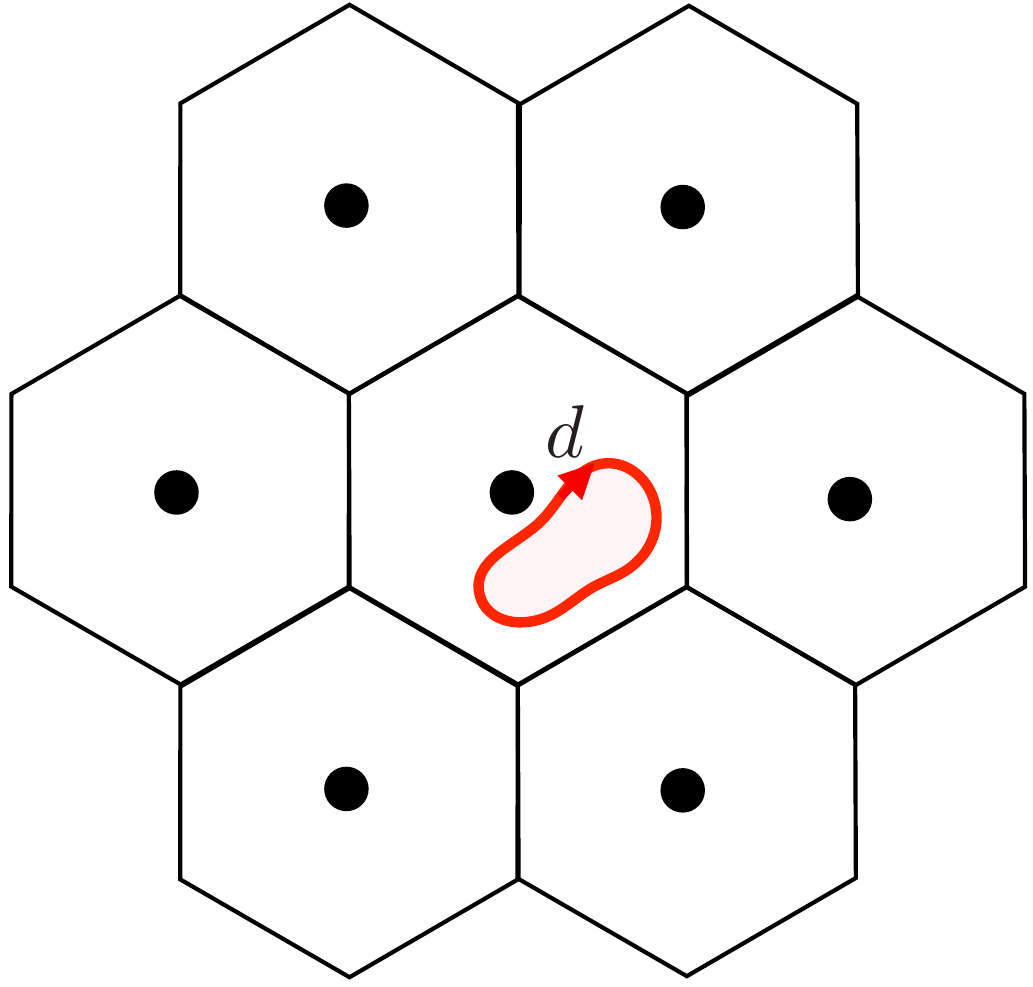}}\\
    &=
    d_{d}\qty(\prod_{f\in S}\frac{\delta^{\bar{d}}_{c_i}}{d_{d}})\    \parbox{4.5cm}{\includegraphics[scale=0.25]{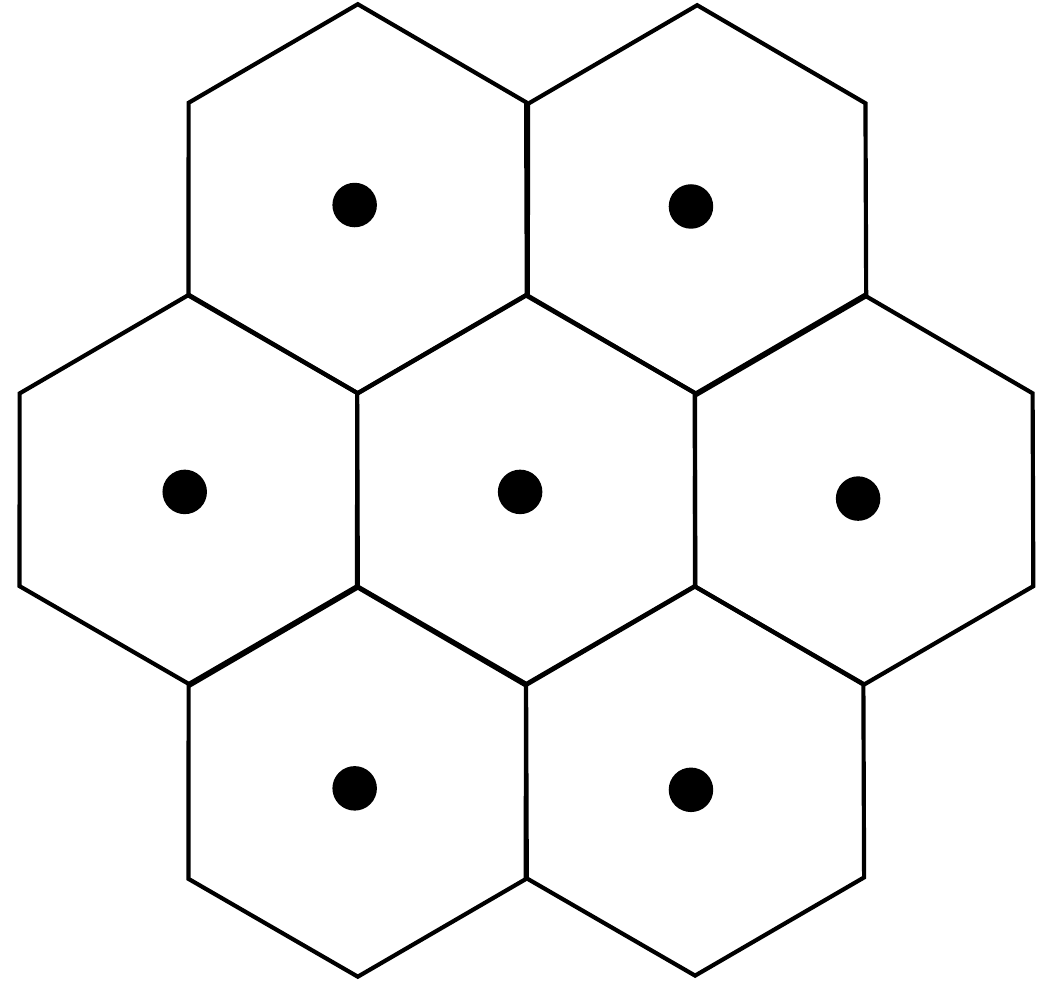}}.
  \end{split}
\end{equation}
By plugging the obtained numerical factor into eq.~\eqref{eq:expvalue_Wilson_loop}, the expectation value of the Wilson loop is (see also ref.~\cite{Ritz-Zwilling:2020itc})
\begin{equation}
\label{eq:wilsonloop_vev}
  \begin{split}
    \expval{\tr U_d(\partial S)}
    &= 
    d_d\prod_{f\in S} \sum_{a_f,b_f,c_f}\psi^*(a_f)\psi(b_f)N_{\bar{a}_fb_f}^{c_f}\frac{\delta_{c_f}^{\bar{d}}}{d_d}\\
    &=
    d_d\qty(\frac{\sum_{a,b}N_{db}^{a}\psi^*(a)\psi(b)}{d_d})^{|S|}\\
    &=d_d\exp(-|S|\sigma_d),
  \end{split}
\end{equation}
where we employ $N_{\bar{a}b}^{\bar{d}}=N_{db}^{a}$ in the second line,
$|S|$ represents the number of the hexagonal plaquettes inside the path $\partial S$, and $\sigma_d$ is the string tension,
\begin{equation}
\label{eq:stringtension}
  \sigma_d\coloneqq\ln\frac{d_d}{\sum_{a,b}N_{db}^{a}\psi^*(a)\psi(b)}.
\end{equation}
If $\sigma_d$ is of order unity, the Wilson loop exhibits the area law.
As is seen later, the wave function in the string net condensed state is $\psi(a)=d_a/\mathcal{D}$~\cite{Levin:2004mi}.
From eq.~\eqref{eq:inverser_dadb}, we find that the string tension $\sigma_d$ vanishes, or equivalently, the expectation value of the Wilson loop in eq.~\eqref{eq:expvalue_Wilson_loop} becomes the unity.

\begin{figure}[t]
\begin{center}
 \includegraphics[width=.5\textwidth]{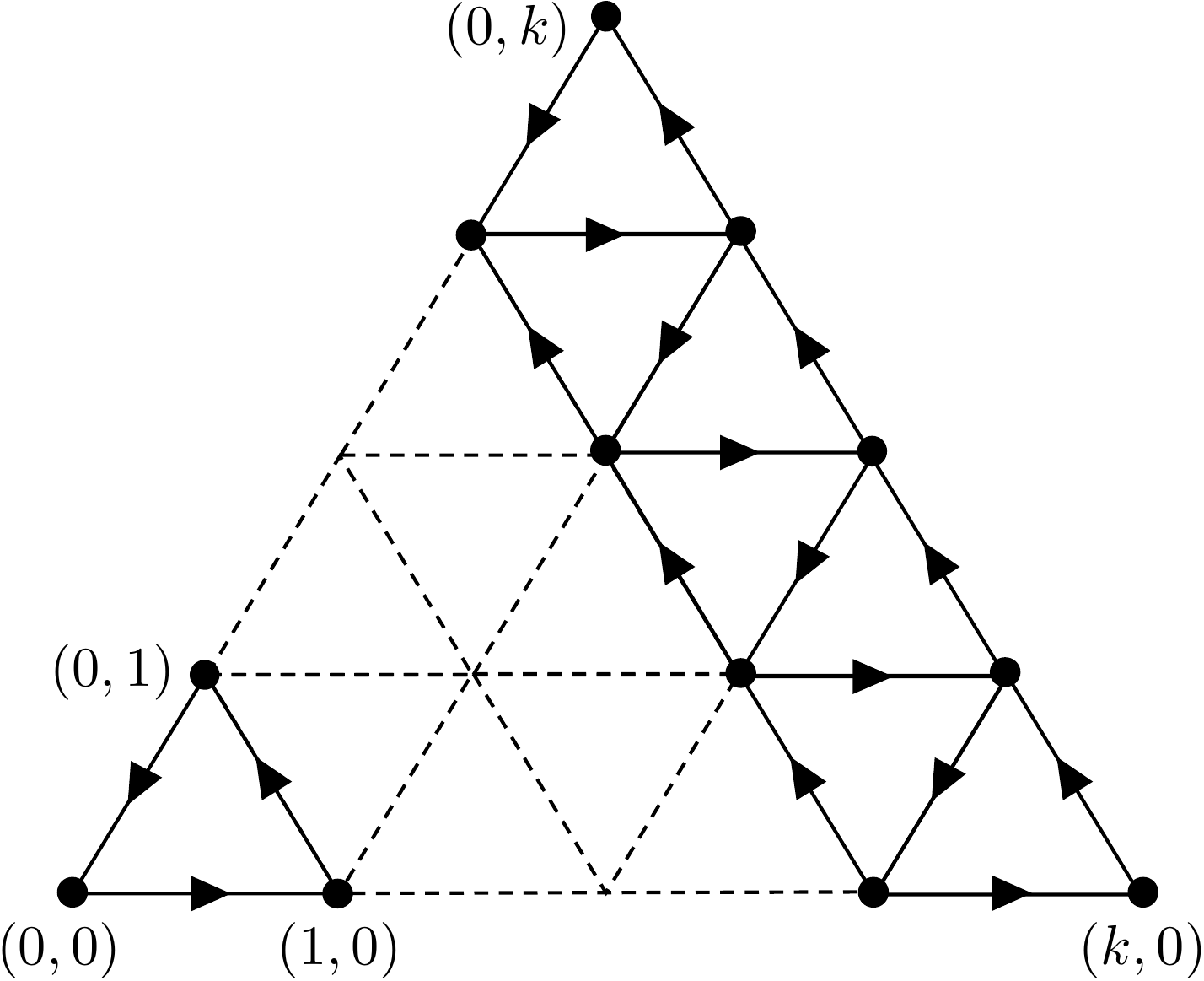}
\caption{\label{fig:boundary}Diagram for $\mathrm{SU}(3)_k$. Arrows indicate the action of the fundamental Wilson loop.}
\end{center}
\end{figure}

Next, we compute the expectation value of the Hamiltonian~\eqref{eq:Hamiltonian}.
Using eq.~\eqref{eq:wilsonloop_vev}, the expectation value of the Wilson loop on the hexagonal plaquette with the representation $(1,0)$ is evaluated as
\begin{equation}\label{eq:exp_trU10}
  \expval{\tr U_{(1,0)}(f)}{\Psi}=
  \sum_{a,b}M_{ab}\psi^*(a)\psi(b),
\end{equation}
where we write $M_{ab}=N_{(1,0)b}^{a}$.
Using the same method to that of the expectation value of the Wilson loop,
we can evaluate $\expval{E_i^2(e)}$:
\begin{equation}
  \begin{split}
    \expval{E_i^2(e)}&=
    \sum_{a,b,a',b'}\psi^*(a')\psi^*(b')\psi(a)\psi(b)
    \bra{\vac} \tr U_{\bar{a}'}(f)\tr U_{\bar{b}'}(f') E_i^2(e)\tr U_{{a}}(f)\tr U_{b}(f')\ket{\vac}.
  \end{split}
\end{equation}
Here, $f$ and $f'$ are the plaquettes adjoining the edge $e$.
The contribution from a plaquette not adjacent to the edge $e$ becomes trivial, as does in the computation of the expectation value of the Wilson loop.
Therefore, it is sufficient to consider the adjacent hexagonal plaquettes.
We use the same technique as before.
First, we apply $E_i^2$ to the state
\begin{equation}
  \tr U_{{a}}(f)\tr U_{b}(f')\ket{\vac}= \ \parbox{4.cm}{\includegraphics[scale=0.3]{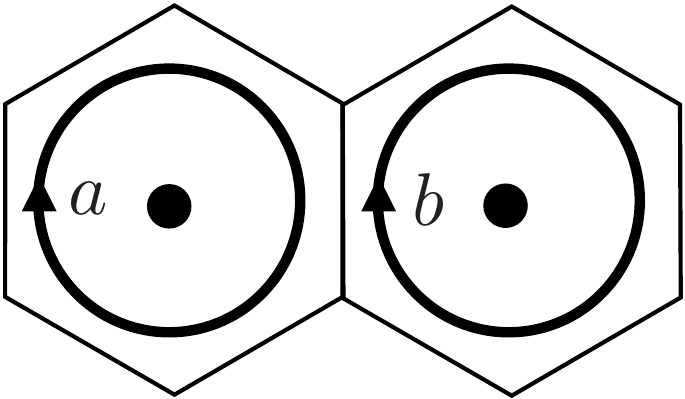}}
\end{equation}
as
\begin{equation}
  \begin{split}
    E_i^2(e) \ \parbox{4.cm}{\includegraphics[scale=0.3]{figs/two_loops1.pdf}}
    &=\sum_{c,\mu}\frac{\sqrt{d_c}}{\sqrt{d_{\bar{a}}d_b}}
    E_i^2(e)\ \parbox{4.0cm}{\includegraphics[scale=0.3]{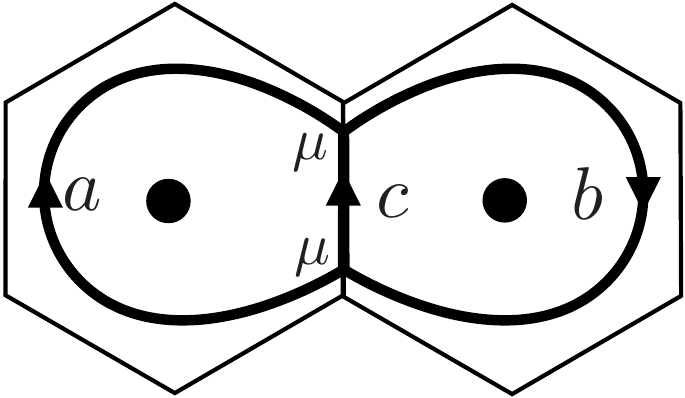}}\\
    &=\sum_{c,\mu}\frac{\sqrt{d_c}}{\sqrt{d_{\bar{a}}d_b}}C_2(c)\
    \parbox{4.0cm}{\includegraphics[scale=0.3]{figs/two_loops2.pdf}}.
  \end{split}
\end{equation}
In the first and second lines, we used eqs.~\eqref{eq:partition} and~\eqref{eq:action_electric_field}, respectively.
Next, we apply the Wilson loop to the state
\begin{equation}
  \begin{split}
    &\tr U_{\bar{a}'}(f)\tr U_{\bar{b}'}(f') E_i^2(e)\tr U_{{a}}(f)\tr U_{b}(f')\ket{\vac}\\
    &=\sum_{c,\mu}\frac{\sqrt{d_c}}{\sqrt{d_{\bar{a}}d_b}}C_2(c)\
    \parbox{4.0cm}{\includegraphics[scale=0.3]{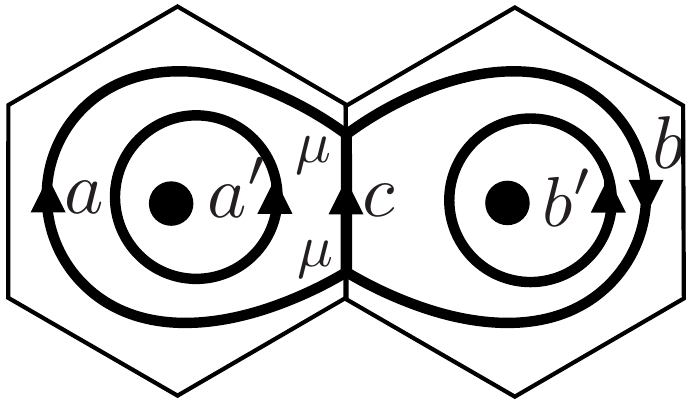}}\\
    &\to\sum_{c,\mu}\frac{\sqrt{d_c}}{\sqrt{d_{\bar{a}}d_b}}C_2(c)
    \frac{\delta_{a}^{a'}\delta_{b}^{b'}}{d_ad_b}\
    \parbox{4.0cm}{\includegraphics[scale=0.3]{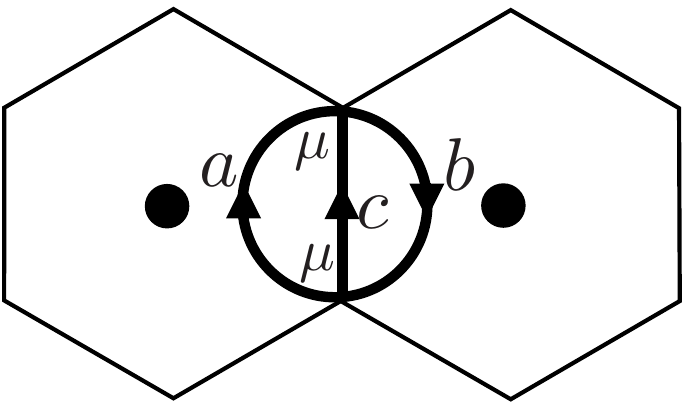}}\\     
    &=\sum_{c}N_{\bar{a}b}^c\frac{\sqrt{d_c}}{\sqrt{d_{\bar{a}}d_b}}C_2(c)\frac{\delta_{a}^{a'}\delta_{b}^{b'}}{d_ad_b}
    \frac{\sqrt{d_{{a}}d_c}}{\sqrt{d_b}}d_b\
    \parbox{4.0cm}{\includegraphics[scale=0.3]{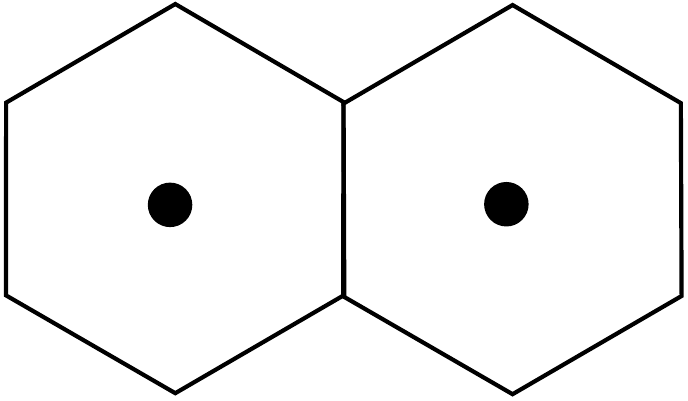}}
  \\ 
    &=\sum_{c}N_{\bar{a}b}^cC_2(c)\frac{d_c}{d_ad_b}\delta_{a}^{a'}\delta_{b}^{b'}\
    \parbox{4.0cm}{\includegraphics[scale=0.3]{figs/two_loops6.pdf}},
  \end{split}
\end{equation}
where we used eq.~\eqref{eq:reduction2} twice to reduce the second line to the third line, eqs.~\eqref{eq:stacking} and~\eqref{eq:loop_deformation} to reduce the third line to the fourth line, and $d_a=d_{\bar{a}}$ to reduce the fourth line to the fifth line.
Therefore the expectation value of $E_i^2(e)$ is given as
\begin{equation}\label{eq:exp_E2}
  \begin{split}
    \expval{E_i^2(e)}&=
    \sum_{a,b,c}C_2(c)N_{\bar{a}b}^c\frac{d_c}{d_ad_b}|\psi(a)|^2|\psi(b)|^2.
  \end{split}
\end{equation}
Finally, let us compute the expectation value of the Hamiltonian.
Using eqs~\eqref{eq:exp_trU10} and \eqref{eq:exp_E2}, the Hamiltonian density reads
\begin{equation}
\begin{split}
  &\calH=\frac{1}{V}\expval{H}
  \\
  &=   \sum_{a,b,c} C_2(c)N^{c}_{\bar{a}b}\frac{d_{c}}{d_{a}d_{b}} |\psi(a)|^2|\psi(b)|^2
- \frac{K}{2}\sum_{a,b}\psi^*(a)\left(M_{ab}+M^\dagger_{ab}\right)\psi(b) ,
  \label{eq:Hamiltonian_one_plaquette_su3}
\end{split}
\end{equation}
where $V$ is the number of the plaquettes, and $M_{ab}$ is the adjacency matrix, which is one between representations connected by arrows in figure~\ref{fig:boundary} otherwise 0.
Note that since each plaquette has four $E_i^2$ terms, and a $E_i^2$ term (an edge) is shared by two plaquettes, the factor of the $E_i^2$ term in eq.~\eqref{eq:Hamiltonian_one_plaquette_su3} is $4\times1/2\times1/2=1$.

\subsection{Minimization of \texorpdfstring{$\mathcal{H}$}{H}}

To minimize $\calH$ with the constraint $\sum_{a} |\psi(a)|^2=1$, we solve the imaginary-time evolution
\begin{equation}
\begin{split}
        \partial_\tau\psi(\tau,a)&=-\sum_{b,c} \qty(C_2(c) N^{c}_{\bar{a}b}\frac{d_{c}}{d_{\bar{a}}d_{b}}|\psi(\tau,b)|^2\psi(\tau,a)+C_2(c) N^{c}_{\bar{b}a}\frac{d_{c}}{d_{a}d_{\bar{b}}}|\psi(\tau,b)|^2)
    \\
    &+ \frac{K}{2}\sum_{a,b}\left(M_{ab}+M^\dagger_{ab}\right)\psi(\tau,b)-\Lambda \left(\sum_b|\psi(\tau,b)|^2-1\right)\psi(\tau,a) ,
\end{split}
\label{eq:tdvp}
\end{equation}
where the last term is a penalty term to impose $\sum_b |\psi(b)|^2=1$.

\begin{figure}[!h]
\begin{center}
 \includegraphics[width=.8\textwidth]{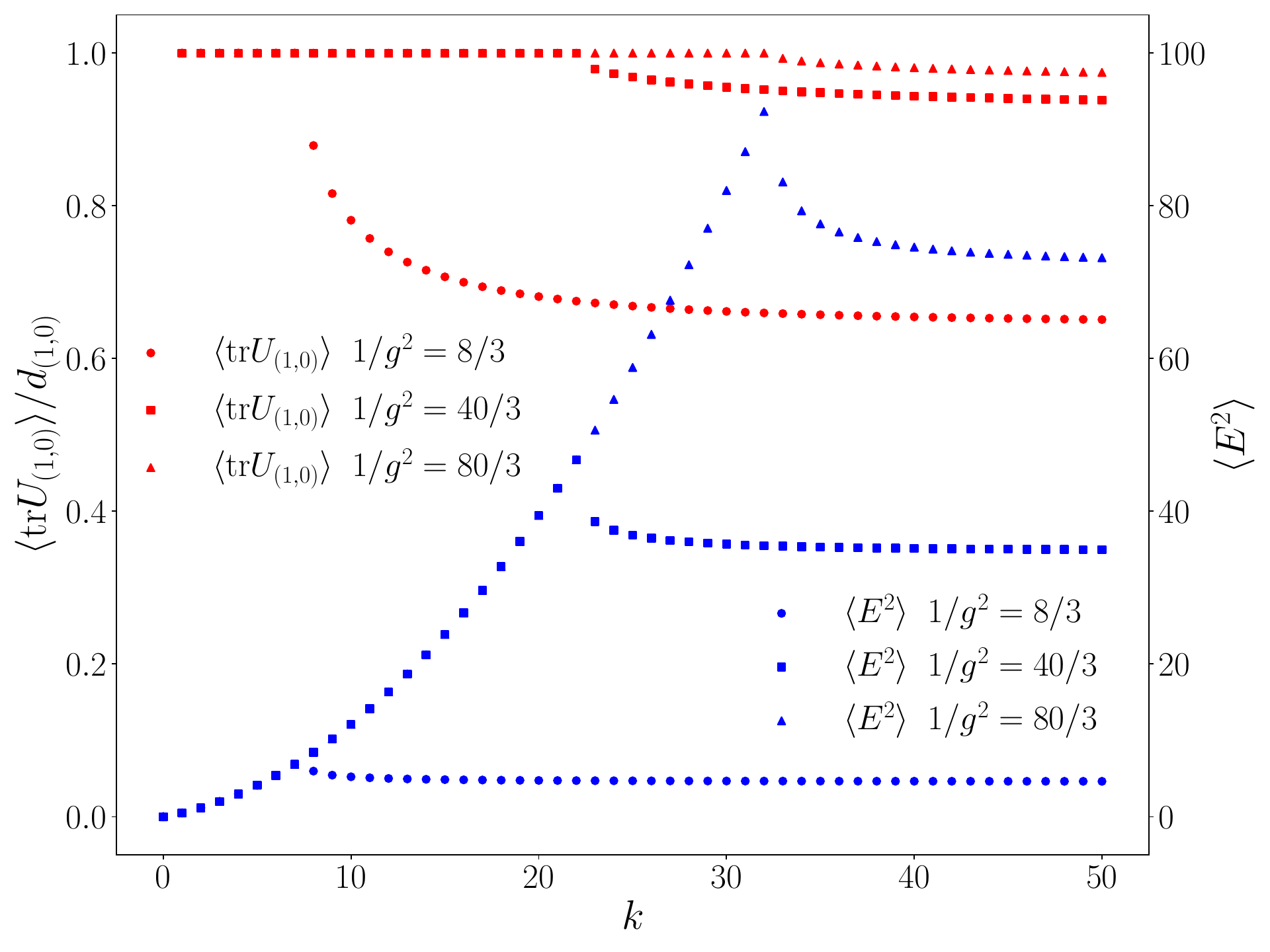}
\caption{\label{fig:kdep}$k$ dependence of the electric and magnetic Hamiltonian density $\langle E_i^2\rangle$ and $\langle {\tr}U_{1,0}\rangle$ with the lattice coupling $1/g^2=8/3$, $40/3$, and $80/3$.}
\end{center}
\end{figure}

\section{Numerical results}
\label{sec:numerical}
We solved eq.~\eqref{eq:tdvp} using the fourth-order Runge-Kutta method.
The initial state is $\psi(0,a)=\delta_{a,(0,0)}$, and we used $\psi(50,a)$ as the groundstate wavefunction.
We take $\Lambda=400\max(K,1)$, and the time step of the Runge-Kutta method  $\delta\tau$ is fixed with $\delta\tau=0.25/\Lambda$. 
First, we show the $k$ dependence of the electric and magnetic Hamiltonian density, $\langle E_i^2\rangle$ and $\langle \tr U_{(1,0)}\rangle$ with changing $1/g^2$ from small to large in figure~\ref{fig:kdep}. 
As is clearly seen in the $k$ dependence of $\langle E_i^2\rangle$, we see the convergence of observables as $k$ increases, and larger $k$ is needed for simulating the groundstate with larger $1/g^2$.
Note that the groundstate is in the topological phase below a critical $k$, where the Wilson loop is topological, i.e., $\langle \tr U_{a}\rangle=d_{a}$ as in eq.~\eqref{eq:loop_deformation}.
We find that a moderate coupling $1/g^2\sim1$ requires $k\sim10$ from figure~\ref{fig:kdep}. This may be reachable in the near future qudit computers~\cite{Ringbauer:2021lhi,Zache:2023dko}, where the basis of a single link or multiplicity is encoded into a single qudit.
Importantly, the convergence accompanies a phase transition. This phase transition is interesting in views of quantum information or condensed matter physics but it is unwanted for simulation of high-energy physics.
As detailed below, $\mathrm{SU(3)}$$_k$ Yang-Mills theory with small $k$ is not smoothly connected to $\mathrm{SU(3)}$ Yang-Mills theory of our target, and thus we should take great care of the $k$ dependence of observables when available $k$ is limited.

\begin{figure}[t]
\begin{center}
 \includegraphics[width=.8\textwidth]{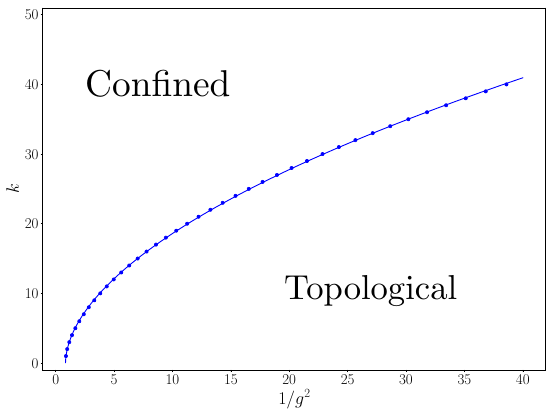}
\caption{\label{fig:phase}Phase structure of $\mathrm{SU}(3)_k$ Yang-Mills theory. The blue dot represents the critical coupling at each $k$. We fitted the data by $k=a(1/g^2-1/g_0^2)^b$, which is shown by the blue curve. The confined phase is smoothly connected to $\mathrm{SU(3)}$ Yang-Mills theory. The topological phase exhibits topological order described by the string-net condensation. The phase transition occurs between these phases when crossing the curve.}
\end{center}
\end{figure}

\begin{figure}[t]
\begin{center}
 \includegraphics[width=.8\textwidth]{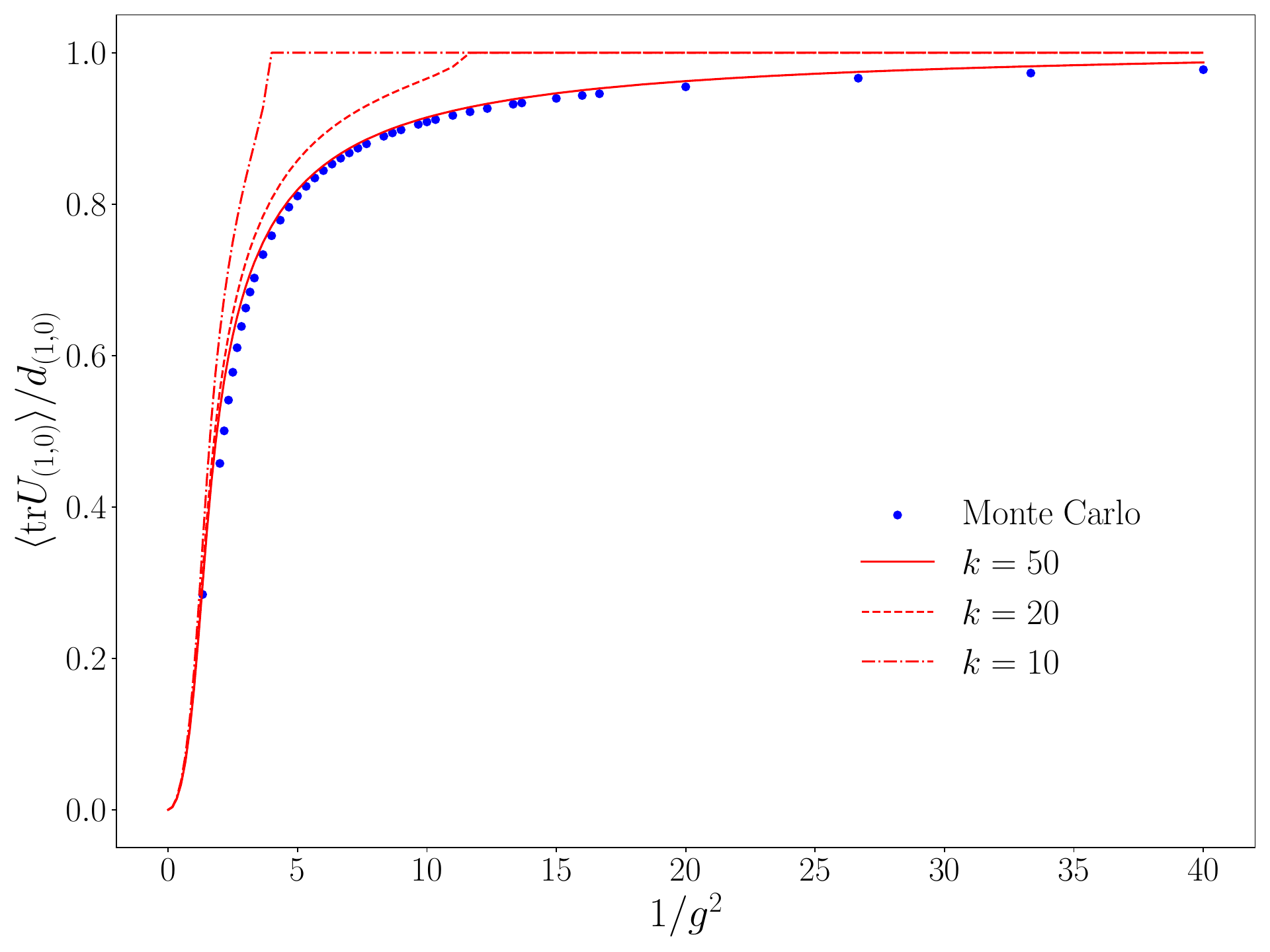}
\caption{\label{fig:MC}$1/g^2$ dependence of the fundamental Wilson loop on the minimum plaquette. Mean field computation of $\mathrm{SU(3)}_k$ Yang-Mills theory (red curves) is compared with the Monte Carlo simulation of continuous $\mathrm{SU(3)}$ Yang-Mills theory (blue dots). Monte Carlo data is taken from ref.~\cite{Bialas:2008rk}.}
\end{center}
\end{figure}

To study the phase transition and phase structure more quantitatively, we fix $k$ and study the $1/g^2$ dependence of observables.
We find that the phase transition occurs as continuously changing $1/g^2$ from small to large between the confined and topological phases (see figure~\ref{fig:MC} for $k=10$ and $20$). We compute the critical $1/g^2$ as a function of $k$, and draw the phase diagram in figure~\ref{fig:phase}. We fitted the data by $k=a(1/g^2-1/g_0^2)^b$, and obtained $a=5.58(3)$, $b=0.543(1)$, and $1/g_0^2=0.838(5)$. The fitting curve is also shown in figure~\ref{fig:phase}. It would be useful for estimating $k$, which is required for simulating the continuum limit.
Now, let us elaborate on the phase structure.
The confined phase is smoothly connected to the vacuum at $1/g^2=0$, i.e., the groundstate in the strong coupling limit, while the topological phase is smoothly connected to the groundstate in the weak coupling limit. 
Within the mean field computation, the groundstate of the topological phase is the same as that of the string-net model~\cite{Levin:2004mi,Zache:2023dko}, i.e., the string-net condensed state $|\Psi_\text{string-net}\rangle=\prod_{f\in \mathcal{F}} U_\mathrm{reg}(f)|0\rangle/\calD^2$, where $U_\text{reg}(f)$ is the Wilson loop of the regular representation, and given explicitly as $U_\mathrm{reg}(f)=\sum_a d_a U_{a}(f)$, and $\calD$ is the total quantum dimension.
The string-net condensed state exhibits topological order, which is classified by the UMTC.
In short, to simulate $\mathrm{SU(3)}$ Yang-Mills theory, which is necessary for high-energy physics, we should keep $k$, so that the state is in the confined phase (an extrapolation to the $k\rightarrow\infty$ limit is also needed), while the state should be in the topological phase to use topological quantum computation or quantum error correcting code~\cite{Koenig:2010uua,Schotte:2020lnz}.

\begin{figure}[t]
\begin{center}
 \includegraphics[width=.8\textwidth]{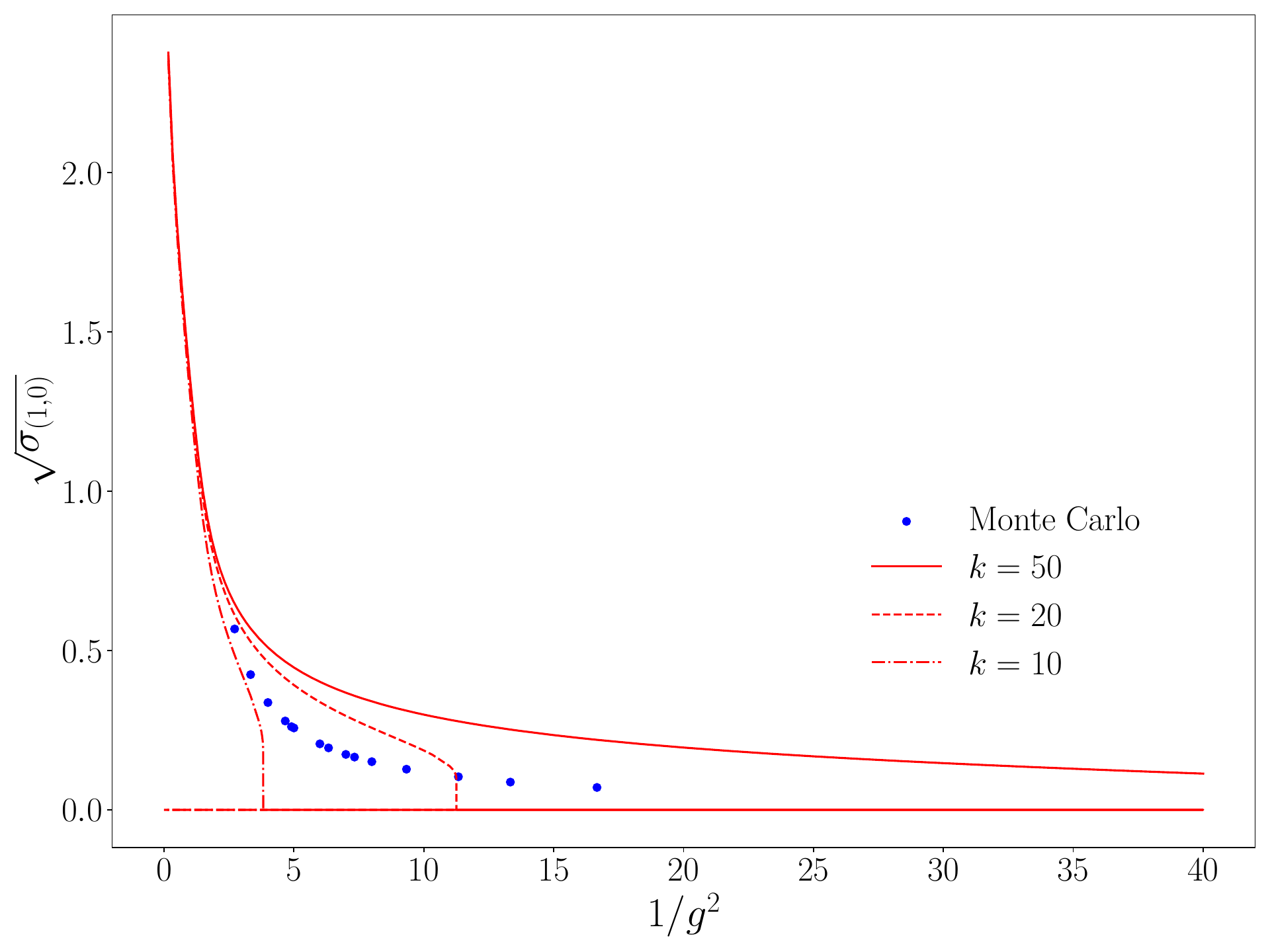}
\caption{\label{fig:tension}$1/g^2$ dependence of the string tension of the fundamental Wilson loop. The mean field computation of $\mathrm{SU(3)}_k$ Yang-Mills theory (red curves) is compared with the Monte Carlo simulation of $\mathrm{SU(3)}$ Yang-Mills theory (blue dots). Monte Carlo data is taken from ref.~\cite{Bialas:2008rk}. We used the lattice unit.}
\end{center}
\end{figure}

\begin{figure}[t]
\begin{center}
 \includegraphics[width=.8\textwidth]{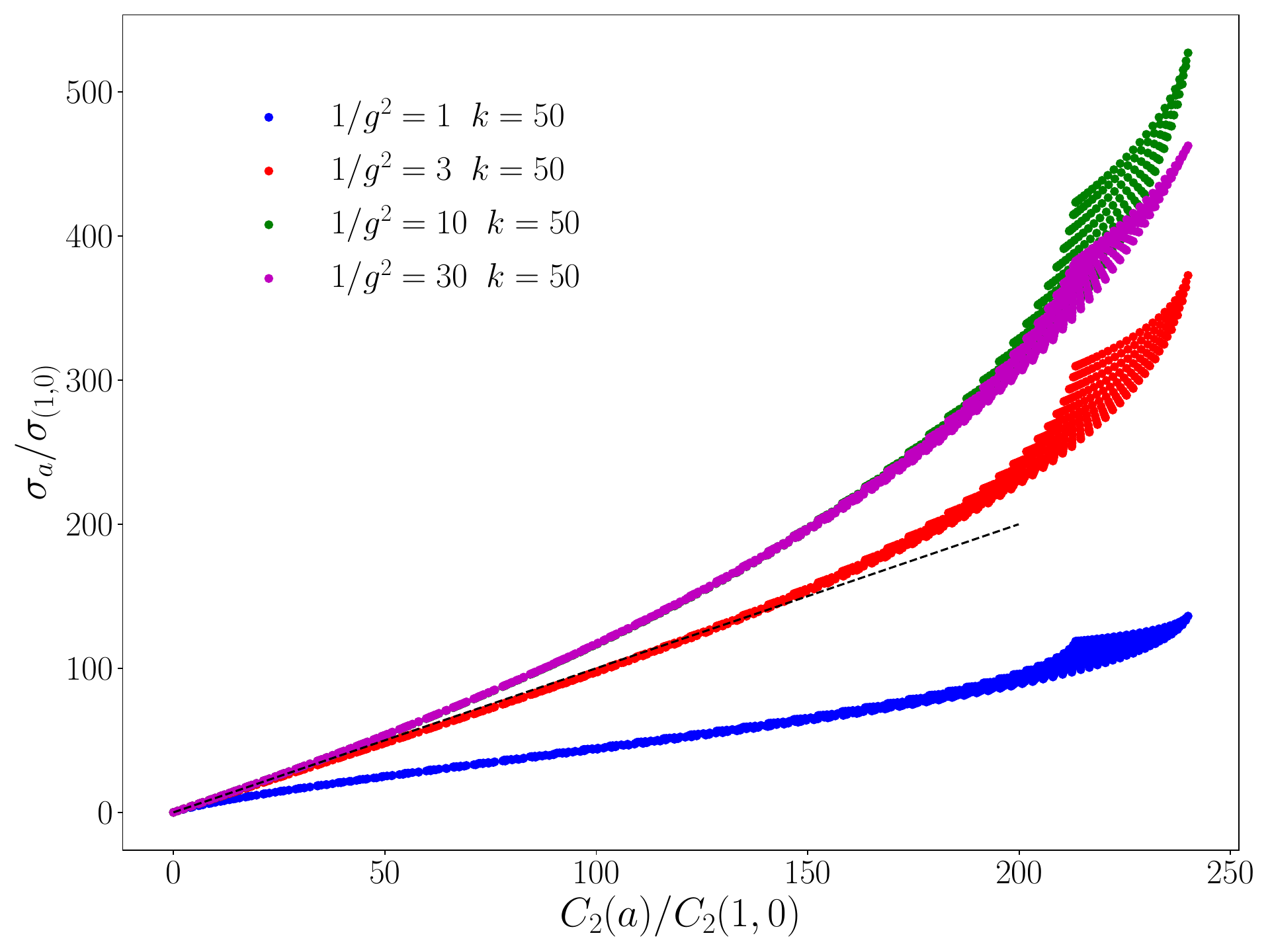}
\caption{\label{fig:casimir}String tension of the Wilson loop of various representations as a function of the Casimir invariant with $k=50$, and $1/g^2=1,3,10,30$. The black dashed line shows the Casimir scaling $\sigma_{a}/\sigma_{(1,0)}=C_2(a)/C_2(1,0)$.}
\end{center}
\end{figure}

\begin{figure}[t]
\begin{center}
 \includegraphics[width=.8\textwidth]{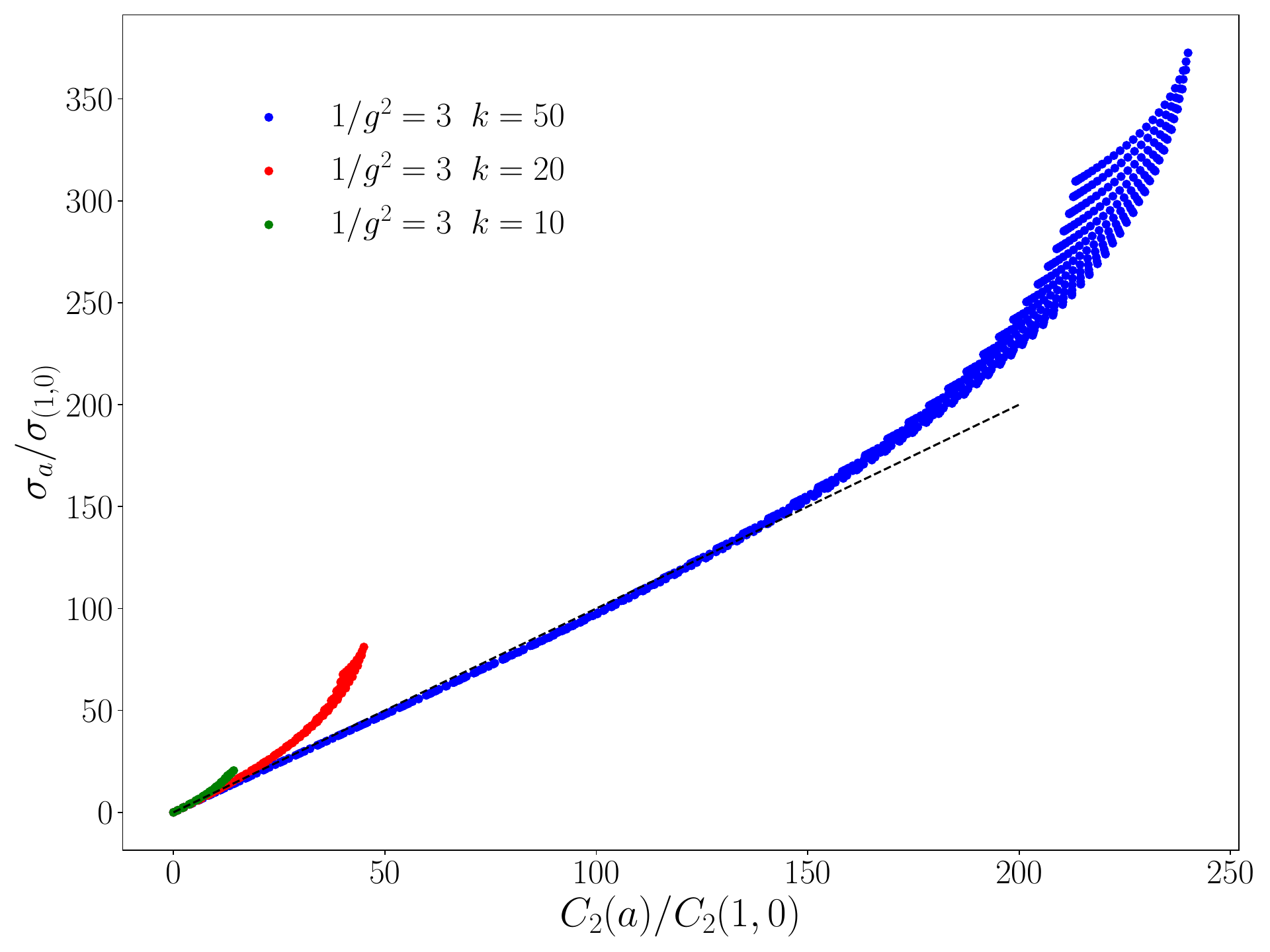}
\caption{\label{fig:casimir_kdep}String tension of the Wilson loop of various representations as a function of the Casimir invariant with $k=10,20,50$, and $1/g^2=3$. The black dashed line shows the Casimir scaling $\sigma_{a}/\sigma_{(1,0)}=C_2(a)/C_2(1,0)$.}
\end{center}
\end{figure}

To give a more quantitative discussion on the physics in the $k\rightarrow\infty$ limit, we compare the mean field computation of $\mathrm{SU(3)}_k$ Yang-Mills theory with the conventional Monte Carlo simulation of $\mathrm{SU(3)}$ Yang-Mills theory. 
We show the expectation value of the fundamental Wilson loop on the hexagonal plaquette in figure~\ref{fig:MC}, where the Monte Carlo data are taken from ref.~\cite{Bialas:2008rk}.
Using sufficiently large $k$, our mean field computation is in good agreement with the Monte Carlo data.
This means that the variational ansatz~\eqref{eq:wavefunction} captures the essential features of Yang-Mills theory.
Note that the number of variational parameters $N_v$ is scaled as $N_v=k^2/2+3k/2+1$, and e.g., $N_v=1326$ when $k=50$.
Furthermore, using eq.~\eqref{eq:stringtension}, we can compute the string tension of the Wilson loop of any representation.
We show the string tension of the fundamental Wilson loop in figure~\ref{fig:tension}.
In this case, we see the quantitative difference between the mean-field and Monte Carlo results even if we use a large $k$.
This may be the fault of the mean-field computation, in which the large Wilson loop is given by the product of the small uncorrelated Wilson loops as described in section~\ref{sec:tdvp}.
It may be rather surprising that such a simple computation leads to quantitative results.
We can generalize the variational ansatz by following refs.~\cite{PhysRevLett.119.070401,Schotte:2019cdg}, which may capture correlations missed in the present ansatz and enable the more correct description of Yang-Mills theory.
Such a generalization may also be important to obtain finite-size corrections to the area law in eq.~\eqref{eq:stringtension}.
Finally, we show the string tension of the Wilson loop of all representations as a function of the second order Casimir invariant with changing $1/g^2$ in figure~\ref{fig:casimir}, and with changing $k$ in figure~\ref{fig:casimir_kdep}.
It is known that the string tension of the Wilson loop of representation $a$ is proportional to the second order Casimir invariant of the representation $C_2(a)$, i.e., $\sigma_a=\kappa C_2(a)$, which is referred to as the Casimir scaling in the literature~\cite{Ambjorn:1984mb,Ambjorn:1984dp,Deldar:1999vi,Bali:2000un}.
From figures~\ref{fig:casimir}, and~\ref{fig:casimir_kdep}, we see that the Casimir scaling holds in the region far from the strong coupling limit, but is still in the confined phase in figure~\ref{fig:phase}.
This is natural because the Casimir scaling does not hold in the strong and weak coupling limits with finite $k$.
In the strong coupling limit, the wave function is $\psi(a)=\delta_a^0$, so that the string tension diverges for all representations. On the other hand in the topological phase, the Wilson loop is always unity, which means the string tension is zero for all representations.


\section{Discussion}
\label{sec:discussion}
We have generalized the formulation of a regularized Hamiltonian for $\mathrm{SU(2)}$ lattice Yang-Mills theory based on the spin network and $q$ deformation to $\mathrm{SU(3)}$.
As a demonstration, we have performed the mean-field computation, which shows good agreements with the Monte Carlo simulation.
The variational ansatz~\eqref{eq:wavefunction} is known to be represented by the tensor network called infinite projected entangled pair states (iPEPS)~\cite{PhysRevB.79.085118,PhysRevB.79.085119,Dusuel:2015sta,PhysRevB.101.085117,Zache:2023dko}. The success of the mean-field computation based on iPEPS indicates that the essential features of Yang-Mills theory can be captured by using tensor networks, so that tensor networks would be useful for future studies of Yang-Mills theory and QCD.

Even within the mean-field approximation, we have several directions to improve the analysis.
First, it is important to compute other observables such as the mass of glueballs. 
We can compute the imaginary-time correlation function of Wilson loops with the present mean-field computation, from which we may read off the lightest glueball mass as is commonly done in the conventional lattice simulations.
Second, we need to generalize the computation to $(3+1)$ dimensions and incorporate fermions for studying QCD.
It would also be important to study the nonequilibrium physics such as thermalization~\cite{Hayata:2020xxm}.
We can study nonequilibrium dynamics by changing the imaginary-time evolution~\eqref{eq:tdvp} to the real-time evolution, and making variational parameters spatially inhomogeneous. Such an analysis may correspond to the time-dependent mean field approximation.
We will address these problems in future works.

\section*{Acknowledgements}
The numerical calculations were carried out on cluster computers at iTHEMS in RIKEN.
This work was supported by JSPS KAKENHI Grant Numbers~21H01007, and 21H01084.

\bibliographystyle{JHEP}
\bibliography{bib}

\end{document}